# The ELFIN Mission

V. Angelopoulos[1], E. Tsai[1], L. Bingley[1], C. Shaffer[1,2], D. L. Turner[1,3], A. Runov[1], W. Li[1,4], J. Liu[1], A. V. Artemyev[1], X.-J. Zhang[1], R. J. Strangeway[1], R. E. Wirz[5], Y. Y. Shprits[1,6], V. A. Sergeev[7], R. P. Caron[1], M. Chung[1,3], P. Cruce[1,8], W. Greer[1], E. Grimes[1], K. Hector[1,9], M. J. Lawson[1], D. Leneman[1], E. V. Masongsong[1], C. L. Russell[1], C. Wilkins[1], D. Hinkley[10], J. B. Blake[10], N. Adair[1,11], M. Allen[1,8], M. Anderson[1,12], M. Arreola-Zamora[1], J. Artinger[1,25], J. Asher[1], D. Branchevsky[1,10], M. R. Capitelli[1,11], R. Castro[1,9], G. Chao[1,13], N. Chung[1,14], M. Cliffe[1,15], K. Colton[1,16], C. Costello[1,17], D. Depe[1,18], B. W. Domae[1,18], S. Eldin[1,18], L. Fitzgibbon[1,2], A. Flemming[1,8], I. Fox[1,5], D. M. Frederick[1,11], A. Gilbert[1,18], A. Gildemeister[1,8], A. Gonzalez[1,15], B. Hesford[1,19], S. Jha[1,17], N. Kang[1,11], J. King[1,17], R. Krieger[1,20], K. Lian[1,8], J. Mao[1,21], E. McKinney[1,22], J. P. Miller[1,17], A. Norris[1], M. Nuesca[1,17], A. Palla[1,17], E. S. Y. Park[1,23], C. E. Pedersen[1,5], Z. Qu[1,5], R. Rozario[1,5], E. Rye[1,18], R. Seaton[1,5], A. Subramanian[1,8], S. R. Sundin[1,2], A. Tan[1,24], W. Turner[1,25], A. J. Villegas[1,25], M. Wasden[1,5], G. Wing[1,17], C. Wong[1,25], E. Xie[1,18], S. Yamamoto[1,5], R. Yap[1,26], A. Zarifian[1,19], G. Y. Zhang[1,27]

1   Earth, Planetary, and Space Sciences Department, and Institute of Geophysics and Planetary Physics, University of California, Los Angeles, CA 90095

2   Currently at Tyvak Nano-Satellite Systems, Inc., Irvine, CA 92618

3   Currently at Johns Hopkins University Applied Physics Laboratory, Laurel, Maryland 20723

4   Currently at Department of Astronomy and Center for Space Physics, Boston University, Boston, MA 02215

5   Mechanical and Aerospace Engineering Department, Henry Samueli School of Engineering, University of California, Los Angeles, CA 90095

6   GFZ German Research Centre for Geosciences, Potsdam, Germany 14473

7   Saint Petersburg State University, St. Petersburg, Russia, 199034

8   Currently at Northrop Grumman Aerospace Systems, Redondo Beach, CA 90278

9   Currently at Raytheon Space and Airborne Systems, El Segundo, CA 90245

10   The Aerospace Corporation, El Segundo, CA 90245

11   Currently at Millenium Space Systems, El Segundo, CA 90245

12   Currently at Aptiv, Agoura Hills, CA 91301

13   Currently at The Boeing Company, Long Beach, CA 90808

14   Currently at SF Motors, Santa Clara, CA 95054

15   Currently at SpaceX, Hawthorne, CA 90250

16   Currently at Planet Labs, Inc., San Francisco, CA 94107

17   Computer Science Department, Henry Samueli School of Engineering, University of California, Los Angeles, CA 90095

18   Electrical and Computer Engineering Department, Henry Samueli School of Engineering, University of California, Los Angeles, CA 90095

19   Currently at Jet Propulsion Laboratory, Pasadena, CA 91109

20   Currently at Mercedes-Benz Research and Development North America, Long Beach, CA 90810

21   Currently at Epic Systems Corporation, Verona, WI 53593

22   Currently at California State Polytechnic University, Pomona, CA 91768

23  Currently at Economics Department, University of California, Los Angeles, CA 90095

24  Currently at Experior Laboratories, Oxnard, CA 93033

25  Physics and Astronomy Department, University of California, Los Angeles, CA 90095

26  Mathematics Department, University of California, Los Angeles, CA 90095

27  Currently at Qualcomm, San Diego, CA 92121

## Abstract

The Electron Loss and Fields Investigation with a Spatio-Temporal Ambiguity-Resolving option (ELFIN-STAR, or heretoforth simply: ELFIN) mission comprises two identical 3-Unit (3U) CubeSats on a polar (~ 93° inclination), nearly circular, low-Earth (~450 km altitude) orbit. Launched on September 15, 2018, ELFIN is expected to have a >2.5 year lifetime. Its primary science objective is to resolve the mechanism of storm-time relativistic electron precipitation, for which electromagnetic ion cyclotron (EMIC) waves are a prime candidate. From its ionospheric vantage point, ELFIN uses its unique pitch-angle-resolving capability to determine whether measured relativistic electron pitch-angle and energy spectra within the loss cone bear the characteristic signatures of scattering by EMIC waves or whether such scattering may be due to other processes. Pairing identical ELFIN satellites with slowly-variable along-track separation allows disambiguation of spatial and temporal evolution of the precipitation over minutes-to-tens-of-minutes timescales, faster than the orbit period of a single low-altitude satellite ($T_{orbit}$~90min). Each satellite carries an energetic particle detector for electrons (EPDE) that measures 50keV to 5MeV electrons with $\Delta E/E$<40% and a fluxgate magnetometer (FGM) on a ~72cm boom that measures magnetic field waves (e.g., EMIC waves) in the range from DC to 5Hz Nyquist (nominally) with <0.3nT/sqrt(Hz) noise at 1Hz. The spinning satellites ($T_{spin}$~3s) are equipped with magnetorquers (air coils) that permit spin-up or -down and reorientation ("reor") maneuvers. Using those, the spin axis is placed normal to the orbit plane (nominally), allowing full pitch-angle resolution twice per spin. An energetic particle detector for ions (EPDI) measures 250keV – 5MeV ions, addressing secondary science. Funded initially by CalSpace and the University Nanosat Program, ELFIN was selected for flight with joint support from NSF and NASA between 2014 and 2018 and launched by the ELaNa XVIII program on a Delta II rocket (with IceSatII as the primary). Mission operations are currently funded by NASA. Working under experienced UCLA mentors, with advice from The Aerospace Corporation and NASA personnel, more than 250 undergraduates have matured the ELFIN implementation strategy; developed the instruments, satellite, and ground systems and operate the two satellites. ELFIN's already high potential for cutting-edge science return is compounded by concurrent equatorial Heliophysics missions (THEMIS, ERG, Van Allen Probes, MMS) and ground stations. ELFIN's integrated data analysis approach, rapid dissemination strategies via the SPace Environment Data Analysis System (SPEDAS), and data coordination with the Heliophysics/Geospace System Observatory (H/GSO) optimize science yield, enabling the widest community benefits. Several storm-time events have already been captured and are presented herein to demonstrate ELFIN's data analysis methods and potential. These form the basis of on-going studies to resolve the primary mission science objective. Broad energy precipitation events, precipitation bands, and microbursts, clearly seen both at dawn and dusk, extend from 10s of keV to >1MeV. This broad energy range of precipitation indicates that multiple waves are providing scattering concurrently. Many observed events show significant backscattered fluxes, which in the past were hard to resolve by equatorial spacecraft or non-pitch-angle-resolving ionospheric missions. These observations suggest that the ionosphere plays a significant role in modifying magnetospheric electron fluxes and wave-particle interactions. Routine data captures starting in February 2020 and lasting for at least another year, approximately the remainder of the mission lifetime, are expected to provide a very rich dataset to address questions even beyond the primary mission science objective.

## 1. Introduction

As we enter the sixth solar cycle since the discovery of the Van Allen radiation belts by Explorer 1 in 1958, recent satellite failures (Space News 2010) attributed to space weather remind us of our ever increasing reliance on space technologies and their associated vulnerabilities. Relativistic ("killer") electrons in the Van Allen belts are responsible for disruption of spacecraft systems (Horne et al., 2013), but can vary significantly during strong magnetic storms, driven by solar wind energy input into the Earth's magnetosphere. The response of the radiation belts to storm-time energy circulation is, however, complex and non-linear: Geomagnetic storms can increase (Fig. 1a), decrease (Fig. 1b), or maintain steady (Fig. 1c) fluxes of relativistic electrons (Reeves et al. 2003, Turner et al. 2019) with probabilities of 53%, 19%, and 28%, respectively, for the three types of response. Such behavior, caused by competition between sources and sinks, requires a careful examination of not only transport and acceleration processes but also of loss processes during storms. NASA, along with a number of other national and international agencies, has executed a multi-billion dollar program of missions such as the Van Allen Probes or Radiation Belt Storm Probes mission (Mauk et al. 2013); ERG or Arase (Miyoshi et al. 2018); and ionospheric suborbital or ground projects, such as BARREL (Millan et al., 2013), StormDARN (Chisham et al. 2007), and incoherent scatter radars (Kelly and Heinselman 2009; Valentic et al. 2013). When coupled with existing assets, such as SDO (Pesnell et al. 2012), ACE (Stone et al. 1998), and THEMIS (Angelopoulos 2008) these deployments create an unprecedented panoply of observatories for studies of radiation belt "killer electron" energization in response to solar variability. This concerted international effort underscores the complexity and the importance of the problem at hand.

Because radiation belt electron fluxes are a balance between sources and losses, it is critical to understand loss rates and mechanisms and their relative efficacy in suppressing equatorial relativistic electron flux levels as a function of storm time. Although precipitation caused by equatorial pitch-angle scattering by EMIC waves is a leading loss candidate mechanism (Thorne and Kennel 1971; Thorne and Andreoli 1980; Bortnik et al. 2006; Millan et al. 2007; Li et al. 2007; Sandanger et al. 2007; Usanova et al. 2008, 2014; Aseev et al. 2017, Xiang et al. 2018;), its effects have yet to be quantified. Another viable mechanism, fast losses to the outer boundary of electron trapping (Nagai et al. 1988, Mathie et al. 2000; Desorgher et al. 2000; Turner et al. 2012b) followed by outward radial diffusion (Brautigam et al. 2000; Shprits et al. 2006a; Turner et al. 2012a), often termed magnetopause shadowing, leads to losses over a broad range of energies and pitch angles beyond a given L-shell (L~4) and over a broad range of MLTs. Of the other potential equatorial scattering mechanisms outside the plasmasphere, chorus wave scattering is the only one that resonates with energetic electrons (e.g., Horne and Thorne 2003; Ni et al. 2008; Thorne et al. 2010; Li et al. 2013). Chorus waves, however, cause pitch-angle diffusion over a broad range of sub-relativistic energies, from tens to a few hundreds of keV. (Hiss waves are limited to the plasmasphere and do not often affect the heart of the outer radiation belt, whereas magnetosonic waves cause energy diffusion and particle acceleration rather than pitch-angle scattering into the loss cone; e.g., Horne et al., 2007; Ma et al., 2016a, 2016b.) As we shall discuss, EMIC wave scattering can be easily differentiated from chorus scattering and magnetopause shadowing and chorus scattering losses by the distinct, energy- and pitch-angle-dependent loss cone signatures EMIC waves create within the loss cone.

High-altitude (e.g., geosynchronous) missions have an inherent disadvantage in studying precipitation. Because the loss cone decreases quickly with altitude (to <10° at >10,000km), distinguishing between temporal variations in trapped flux and enhancements in precipitation (let alone spectral details within the loss cone) is technologically demanding (Kasahara et al., 2018) if not impossible at such altitudes. Reasonable pitch-angle resolution in the loss cone (2-3 points over 30°-60°) at altitudes 350-1,000km, however, is easily achievable with heritage particle instruments, enabling measurements of both trapped and precipitating particle fluxes and adequate characterization of loss mechanisms from such a vantage point. The primary objective of ELFIN is to understand active-time precipitation of radiation belt relativistic electrons in particular, as it pertains to action by EMIC waves.

By directly measuring precipitation rates as a function of energy and latitude at least once approximately per 2 hours (an orbit period) during the course of storms, ELFIN is designed to determine whether main phase electron dropouts are produced by EMIC scattering and related atmospheric losses rather than other mechanisms (such as chorus waves or magnetopause losses). ELFIN will be the first mission to accurately measure the energy and pitch-angle spectrograms of relativistic precipitating particles, and by comparing those measurements with models of EMIC wave scattering, to determine whether such waves are responsible for electron scattering losses. By comparing ELFIN's electron loss measurements with equatorial measurements of electron transport and acceleration from available missions (such as THEMIS, the Van Allen Probes, and ERG), the ELFIN team will contribute significantly to tour understanding of space weather.

ELFIN's observations come on the heels of a very active period of radiation belt studies aided by the operation of the Van Allen Probes (2012-2019). Amongst past, polar-orbiting missions, SAMPEX has provided some of the most detailed, long-duration, and important information on radiation belt losses to-\ date. From a 600 km altitude, an ~82° inclination orbit, and using broad integral energy bins, it measured precipitating electrons at >1MeV and trapped/quasi-trapped electrons of 2-6MeV energy (Baker et al. 1993).

Relativistic electron precipitation structures are categorized (e.g., Nakamura et al., 2000, see their Fig. 1) as "broad precipitation" covering the entire outer radiation belt (L-shells between ~3 and 6), "precipitation bands" (PBs) across $\Delta L \sim 0.5$-1 (see also: Blum et al., 2013), and "microbursts" across $\Delta L \sim 0.01$-0.05 (0.3-2sec on a polar orbiting, low-altitude satellite; see also: O'Brien et al. 2004).

An example of precipitation bands is shown in Fig. 2a (from Bortnik et al. 2006, Fig. 3). Later, balloon-based X-ray observations of similar duration and at similar latitudes were also identified as caused by precipitation bands (Millan et al. 2002). Such bands were attributed to EMIC waves scattering at the equator (because of their high energy), even though neither the precipitating electron energy and pitch-angle spectrum nor direct measurements of EMIC waves at the ionosphere, let alone the equator, were available. SAMPEX also observed discrete "microburst" precipitation structures (Fig. 2b) associated with *discrete chorus* elements at the equator (Bortnik et al. 2006), although at relativistic electron energies, EMIC waves were also thought to play a significant role. Since then, attempts to infer the relative importance of the two types of loss – EMIC or chorus related - remain inconclusive. One of the issues is that although MEPED detector data from the NOAA/POES satellite, which have crude pitch-angle resolution (separating with a broad field-of-view the down-going from the trapped fluxes) have been found to often have contamination from ring current ions prohibiting accurate computation of loss rates and distinction between loss mechanisms (Lorentzen et al., 2001; Sandanger et al. 2009; Horne et al. 2009; Lam et al. 2010). After methods were devised to clean up the hundreds of keV proton contamination, a statistical study of POES relativistic electron precipitation (Yahnin et al. 2016) showed that their scattering loss correlates well with tens of keV *proton* precipitation, consistent with EMIC waves. However, their loss is also occasionally correlated with lower energy (30- 300keV) *electron* precipitation likely caused by chorus or hiss. Modeling of EMIC wave properties using quasilinear theory (e.g., Ni et al. 2015; Capannolo et al. 2019) has shown that for typical wave characteristics (frequency, propagation angle, and amplitudes), $H^+$ and $He^+$ EMIC waves are the most likely potential contributors to ~1-2 MeV electron losses, whereas O+ waves can be implicated in losses at higher energies.

Douma et al. (2018) studied SAMPEX *data* in conjunction with ground-based observations of waves. They found that microbursts, thought to be a potentially very significant relativistic electron loss process (O'Brien et al. 2004; Greeley et al. 2019), are statistically most likely to be associated with whistler mode waves (chorus), rather than EMIC waves, although EMIC wave activity is still occasionally correlated with such precipitation. Using data from the FIREBIRD II CubeSats, Crew et al. (2016), showed that microbursts often extend from 1MeV down to 200keV, which suggests that either EMIC wave theory must be modified to lower the resonant energy to <400keV, or chorus or other waves are primarily responsible for the relativistic electron precipitation. Using conjugate Van Allen Probes and FIREBIRD II measurements, Capannolo et al. (2019) demonstrated that EMIC waves are associated with energetic electron precipitation, at least down to ~250 keV.

Breneman et al. (2015) analyzed BARREL and Van Allen Probe data, demonstrating excellent coherence of relativistic electron precipitation, inferred from X-ray spectra, with whistler mode hiss waves on a global scale and at time-scales as short as minutes. As chorus waves modulated by ULF pulsations or nightside injections seed hiss in the plasmasphere (Bortnik et al. 2008), this suggests that chorus waves not only scatter relativistic electrons directly and locally outside the plasmasphere, but also induce precipitation indirectly on a global scale through powering hiss activity within the plasmasphere.

Analysis of riometer absorption (a proxy for electron precipitation) and POES data together with ground-based observations of EMIC waves provide good evidence for operation of electron scattering by EMIC waves at energies as low as 300keV (Clilverd et al., 2015). The good correlation between the structure and modulation of X-ray-inferred electron precipitation fluxes at BARREL with equatorial observations of EMIC waves provide further good observational support for EMIC scattering of relativistic electrons (Blum et al., 2015). And, indeed, BARREL observations of inferred electron spectra have shown distinctly peaked spectra around 1.0-1.2 MeV that are also consistent with EMIC waves (Clilverd et al., 2017).

Using in situ equatorial measurements, Usanova et al. (2014) showed a good correlation between near-field-aligned pitch-angle flux depletions at the Van Allen Probes and EMIC waves on the ground. The correlation was evident on timescales of hours (the revisit time of the same L-shell by a spacecraft). In an event analysis, Bingley et al., (2019) reported characteristic "bite-outs" in pitch-angle spectra (depletions in near-field-aligned flux) with resonant energy and pitch-angle that matched (spectrum-after-spectrum) the expectation from linear theory using EMIC waves measured simultaneously on the same observational platform, the Van Allen Probes. In addition, flux minima at around 1MeV energy at L~4 were identified in consecutive radial phase space density profiles when EMIC waves were present (Shprits et al., 2017; Aseev et al., 2017), which also bespeaks EMIC losses. The aforementioned observations from the Van Allen Probes suggest that EMIC waves leave indelible marks on the relativistic electron pitch-angle and radial phase space density distributions near the equator. Even more direct evidence of loss by EMIC waves has been reported in Zhang et al. (2016): Using differences in phase space density on consecutive orbit tracks of Van Allen Probe data they showed that EMIC wave detection was well-correlated with depletions (losses) of near-field-aligned fluxes. However, because all the aforementioned equatorial observations do not resolve the loss cone but are made principally outside it, compelling evidence of precipitation in good agreement with observations of EMIC waves and their characteristics (frequency, amplitude, wavenumber) is still lacking.

Electron pitch-angle scattering simulations (Li et al., 2007; Summers and Thorne, 2003; Albert, 2003; Shprits et al., 2006b) show that equatorial EMIC waves, which can produce significant loss of electrons at energies as low as 0.4 MeV (see Fig. 3, for 1MeV electrons), are far more effective than chorus waves in reducing relativistic electron lifetimes. Such simulations result in predictions of characteristic loss-cone energy and pitch-angle spectra by EMIC waves that can be used to distinguish this scattering mechanism from others using particle observations in the loss cone alone. Pitch-angle distributions as a function of energy at fixed energy provide a telltale signature (Fig. 3, bottom) that can be used to determine whether electrons of ~0.5-2 MeV are scattered by EMIC waves. As we shall see in Section 2.4, energy and pitch-angle spectra as a function of time since the waves' appearance (and as function of candidate waves) can be used to determine the minimum resonant energy of precipitating electrons. When measured *in situ* by ELFIN, or at ground stations or equatorial satellites conjugate to ELFIN, EMIC waves can also be compared with the findings from ELFIN's spectra to determine whether they are in agreement with the proposed scattering mechanism. Thus, for the first time, ELFIN can obtain energy- and pitch-angle spectra of precipitating relativistic electrons with sufficient resolution to determine (from consecutive radiation belt crossings) not only electron loss rates but also whether equatorial EMIC waves are responsible for electron scattering or whether other mechanisms are more important in depleting storm-time radiation belt fluxes.

ELFIN's mission concept (Fig. 4) was developed at UCLA in 2008 to be a scientific front-runner amongst several competing proposal ideas for internal support and for seeking a competitive selection mechanism to establish a hands-on student educational program. FGM electronics miniaturization, accomplished under NASA Planetary Instrument Definition and Development Program (PIDDP) funds and further subsidized by internal funding, resulted in the multi-chip module (MCM) implementation, as flown. Energetic particle detector (EPD) development, with participation and advice from T. A. Farley (the developer of UCLA's OGO-5 particle detector, see: Walker and Farley, 1972) resulted in a re-design of the shielding and anti-coincidence logic based on experience from the THEMIS SST instrument, and was geared specifically towards the radiation belt environment. Significant effort was expended to ensure a very low level of background (verified through modeling with realistic environment fluxes) and, in addition, through development and testing of a coincidence electronics chain (see Section© 3). The first incarnation of the FGM and EPD instrument suite was delivered on the Lomonosov spacecraft in July 2011, under separate NSF and internal UCLA funding (Shprits et al., 2018). This polar-orbiting Russian mission was launched in the late Fall of 2013, provided the first measurements of electron precipitation spectra from EPD, albeit without the pitch-angle information of the ELFIN CubeSat. Development continued between 2011 and 2015 under the US AirForce's University Nanosat Program (UNP-8 selection) and UCLA internal funding. This allowed further development of the instrument suite's flight software, building a thermal vacuum (TVAC) facility for CubeSat testing and the UCLA ground antenna for CubeSat up/downlinks compatible with the Global Educational Network for Satellite Operations (GENSO) network, seeking eventual selection for flight by NSF or NASA. In 2014, both NSF and NASA selected ELFIN (as a single-satellite mission) for development with joint funding. At that time, long-lead items were purchased and an undergraduate student team of ~40 students was selected and was structured hierarchically into various subsystems (with staff mentorship) using a NASA-style project organization and management process. Traditional reviews (Science Requirements Review and Systems Requirements Review) set off the program's full-blown development. UCLA's major partner, The Aerospace Corporation, also joined the program shortly thereafter, providing review support, avionics boards, and advice on programming and other subsystems. In 2014 NASA's CubeSat Launch Initiative (CLSI) program also selected ELFIN as one of its missions, committing to secure an acceptable ride to polar orbit under its Educational Launch of Nanosatellites' 18[th] flight opportunity (ELaNa XVIII). The IceSatII launch opportunity was the secondary ride option identified for ELFIN. It was initially slated for launch in the summer of 2017. However, a one-year launch delay, announced in Fall 2016, entailed cost increases to the ELFIN program and prompted a proposal to NASA to execute a provision in its 2014 selection for an ELFIN science enhancement. (That provision stated that should cost and schedule resources became available during the program a two-satellite ELFIN mission would be considered for funding.) The revised mission was named ELFIN with Spatio-Temporal Ambiguity Resolution (ELFIN-STAR or initially ELFIN*) but henceforth is called just ELFIN, for simplicity. Given that sufficient time had now become available to order additional detectors, radios, and other long-lead items, the second satellite-build as proposed by the ELFIN team was approved and funded by NASA by November 2017. It enabled delivery of a second flight unit, together with the first, in the summer of 2018, in time for the ELaNa XVIII launch. Fortuitously, a second spot on ELaNa XVIII became available around the time of the second ELFIN satellite's selection. After a second CLSI proposal by the ELFIN team, for the second CubeSat's ride to space, that CubeSat was also selected to obtain a spot on the same launch vehicle as the first one, enabling the joint launch of the two ELFIN satellites to space.

At the time of its 2014 selection by NSF and NASA, ELFIN's predecessor, the zenith-only viewing instrument suite on the non-spinning Lomonosov spacecraft, ELFIN-L, had already demonstrated the successful in-flight operation of the first incarnation of the instruments. Thus with ELFIN's selection by NSF and NASA, the team ramped up its spacecraft development efforts with additional help from The Aerospace Corporation's successful, on-going CubeSat development program (capitalizing on lessons from AeroCubes 1-8, MEPSI, and PSSC). ELFIN's instrumentation maturity allowed us to make iterative improvements to the instrument suite (e.g., reducing electronics noise) and to better focus on bus development, integration and testing (I&T). Even so, significant hurdles associated with the compact CubeSat design, such as spacecraft electronics noise mitigation, high launch loads, tight tolerance requirements (as low as 2 thousands of an inch), and solid state detector delivery schedule, detector integrity and quality, had to be overcome. Some of these, discussed further in Tsai et al. (2020), have affected the final mission but not enough to threaten prime mission success. One of the most important legacies of the mission is the professional training of hundreds of students (>250 thus far) and many early-career scientists and engineers over the last several years. The challenges encountered have made a strong, positive impact on their professional development. The success of the mission is a testament to their perseverance. Now in either industry or graduate school, they often comment on the transformative effect the project has had on their lives due to its high pace of development; realistic challenges; team diversity; and freedom to fail and try, try again. ELFIN is projected to last in orbit through at least February 2021.

The EPDE has now been commissioned and is calibrated, allowing us to fully address the prime mission objective. The FGM is operating nominally and at low noise. EMIC wave detection, however, requires further calibration using boom-to-body rotation matrices that have yet to be developed. Continuing data collection from the FGM will enable future wave detection by proper analysis; the processing software to perform such high-quality rotations is still under construction. The EPDI has been operated in orbit. Its electronics are functioning nominally. Because additional layers of epoxy were inadvertently placed on the active area by the detector provider (Micron) during a process-change in that facility, the inactive detector thickness is ~1.5microns (rather than the 0.1 micron thin-window specified). This thickness prevents <250keV protons from being detected. Even though the geometric factor of the EPDI (optimized for 50keV ions to address the secondary mission objective) results in low counts at energies >250keV, the instrument background rate (<1c/s) is sufficiently low that summing over multiple spins and pitch angles may allow us to achieve part of the secondary science objective proposed for energies >250keV. Regular operations over the remainder of the mission commenced in February 2020. These operations, which are expected to result in a year's worth of high-cadence, high-quality data collections and science return, are expected to fulfill most if not all mission objectives.

The second ELFIN satellite, ELFIN-B, is also performing nominally, following the first (ELFIN-A) by an along-track separation of 6500km (~15min) in November 2019. The along-track velocity separation, $\Delta V_{Nov2019} \sim 0.18$ m/s ($V_{A,B} \sim 7.63$km/s), has been increasing since late October 2018 ($\Delta V_{Oct2018} \sim 0.06$ m/s) due to orbit decay. A smaller initial separation velocity was imparted by the launch vehicle ($\Delta V_{Sep2018} \sim 0.03$ m/s) as ELFIN's request for a large inter-satellite velocity separation vector ($\Delta V_{requested@launch} > 1$ m/s) could not be accommodated due to the lateness of the second ELFIN satellite's approval for flight. In the first 2 weeks of October 2018 (during the commissioning phase) the separation velocity was doubled (from 0.03m/s to 0.06m/s) by differential drag. This doubling was thanks to an on-orbit experiment that entailed deploying ELFIN-A's FGM boom but not ELFIN-B's for a period of 14 days. This established a proof of principle that differential drag can be successfully used to control separation. Differential velocity will be further increased later in the mission by orbit decay and an increase in the relative atmospheric drag at lower altitudes. However, this increase will be further aided by attitude maneuvers (such as by placing the spin-axis on the mean orbit plane). This will give ELFIN the opportunity to explore a large range of scales for a short time period near the end of the mission.

Data from several storms have already been collected in orbit, resulting in significant science return. Some of these data will be presented here to demonstrate the mission's capability to obtain high-quality science, sufficient for primary mission closure, as well as its promise for additional science return during the remainder of its lifetime, given the anticipated higher data volumes. The mission objectives, including a traceability matrix and compliance with it, are discussed in Section 2. Sections 3 and 4, discussing ELFIN's instrumentation and bus design, respectively, demonstrate how these mission elements satisfy the proposed science objectives at a top level. A more in-depth presentation of the project implementation covering the project design, development, verification, launch, and in-flight operations can be found in Tsai et al. (2020). The EPD design and implementation are detailed in Caron et al. (2020), and the EPD's in-flight performance and calibration are discussed in Wilkins et al. (2020). The FGM's design and implementation are detailed in Strangeway et al. (2020). The ground system design philosophy, the ELFIN ground antenna, and the space-to-ground communications are discussed in Section 5. An overview of ELFIN's mission and science operations is presented in Section 6. More in-depth discussion of these two sections appears in Palla et al. (2020). Section 7 presents data collected from a recent storm to demonstrate the data quality and analysis procedures and to show-case the promise of the mission to address its science objectives.

## 2. Science Objectives

The primary objective of ELFIN is to elucidate storm-time precipitation of radiation belt relativistic electrons by EMIC waves by determining the storm-time electron precipitation rate and whether the electron pitch angle and energy spectra near the loss cone are consistent with EMIC wave scattering, or other scattering mechanisms. No other mission (past or present) has or will be able to accomplish this goal. For example, the recent Van Allen Probes mission and the currently operational ERG and THEMIS missions are near- equatorial. Although optimal for evaluating acceleration and dynamics of trapped electrons, those missions alone cannot distinguish between precipitation and other mechanisms (drift, diffusion, shadowing) to explain losses. Because the loss cone is no larger than a few degrees at the equator, depending on L-shell, its spectral details cannot be easily resolved there. Past/current low-altitude missions did/do not have the pitch-angle resolution (e.g., SAMPEX and Firebird II had/have angle resolution ~60º and ~45º, respectively) or sufficient residence time at low altitudes (POLAR, Cluster were designed as highly eccentric, high-altitude missions), or provide only a single measurement point within the loss cone and were/are thus unable to determine loss-cone characteristics (NOAA/POES). Like SAMPEX, CubeSat CSSWE (Li et al. 2012) does not have the pitch-angle resolution required to address ELFIN's objective. ELFIN's instrumentation, attitude control, and data capture strategy have been optimized to achieve closure on its primary objective. By comparing the measured loss-cone spectra with modeled spectra of EMIC wave scattering, the ELFIN team will be able to assess whether EMIC waves are responsible for the observed electron losses. Although prime science will be conducted at storm times, ELFIN's particle and magnetic field instrumentation will benefit magnetospheric science at non-storm times, as well.

Additionally, the second ELFIN satellite allows us to go significantly beyond the aforementioned single-satellite objective. Using two satellites, ELFIN will disambiguate the temporal variation of precipitation from its spatial evolution. By determining the temporal change of precipitating flux and the loss-cone spectra on time-scales consistent with EMIC diffusion rates (minutes to several tens of minutes, i.e., much shorter than a low-altitude polar orbit period of ~90min), ELFIN will quantitatively determine the global effects of EMIC waves in depleting radiation belt fluxes. This is possible thanks to the identical build and common launch of the two ELFIN satellites, such that the pair of satellites can perform coordinated, along-track, time-separated observations of the same precipitation region. Such differential measurements enable accurate determination of truly temporal variations, which permit ELFIN to determine whether the rate of flux increase within the loss cone (the loss-cone fill rate) matches the expected rate (and evolution) of scattering using realistic EMIC wave properties at the equator.

The secondary objective of ELFIN is to infer the location of field-aligned current (FAC) sources relative to the physical boundaries of plasma populations during non-storm times, advancing our understanding of processes related to the generation of those currents. ELFIN will be able to determine the rate of change of the current intensity and latitude (e.g., poleward expansion) as a function of time during the course of substorms. The correlation between isotropy boundary location/motion and FAC location/motion is a significant aspect of where and how magnetospheric energy is dissipated to the ionosphere during the late expansion/early recovery of a substorm. Secondary science is discussed further in Section 2.2. Bonus science, enabled by ELFIN's synergy with other missions, is presented in Section 2.3.

### 2.1 Primary Science Objective and Measurement/Orbit Requirements

ELFIN's instrumentation, mission design, and data capture strategy have been optimized for characterizing precipitating electron properties for at least one storm (Table 1). Its primary science is accomplished by obtaining several latitudinal profiles of relativistic electron pitch angle and energy distributions through a storm main/recovery phase, at a ~90min cadence (nominal) in MLTs where EMIC waves are most frequent (post-noon/dusk/pre-midnight sector). Loss rates can thus be determined and electron spectral properties can be used to infer whether EMIC waves are responsible for scattering, or whether other waves or processes are responsible for the observed trapped particle flux reduction as a function of storm time. Note that EMIC wave presence at low altitudes is neither a necessary nor a sufficient condition for science closure: It represents additional validation of the EMIC scattering hypothesis, and it is part of the baseline (though not of the minimum) mission requirements.

The minimum requirements can be achieved by a unidirectional electron detector that provides loss-cone spectral details from a spinning platform (pitch-angle resolution<30°, energy resolution $\delta E/E\sim40\%$) plus an attitude magnetometer that provides the vehicle attitude and spin-phase when its measurements are compared with Earth's magnetic field model (such as the International Geomagnetic Reference Model). Attitude knowledge accuracy requirements are apportioned to magnetic field knowledge (<2°, a fairly relaxed requirements) and particle FOV (<28°, consistent with EPD design). A minimum mission duration requirement of three months was initially obtained assuming a nominal recurrence rate of one storm per month, which is typical of solar maximum or the declining phase of the solar cycle. However, although ELFIN was launched and operates in solar minimum, its long mission lifetime has already permitted three storm-time captures during the first year of operations; most of the data from those have already been downlinked. The mission's anticipated lifetime, approximately 2.5 years, allows far greater science yield than originally anticipated.

Given ELFIN's initial altitude of ~450km (Table 2) and expected decay to <350km after 2.5 years, its average altitude is expected to be ~400km. Projected to that altitude, the equatorial loss cone at L=4 (corresponding to the heart of the storm-time outer radiation belt) maps to a range of pitch-angles <65°, in the north, or greater than their supplementary in the south (Fig. 5). The trapped population can always be sampled by ELFIN with at least one sector, near 90°, given EPDE's angular resolution of <±11.25° (FWHM <22.5°). Below 350km the trapped population cannot be uniquely sampled, but mission duration at that altitude is negligible. To resolve the loss cone, the B-field must be near the spin-plane with a ±15° tolerance (Table 1), a reasonable compromise between detector width, dipole rocking during the course of the day, and loss-cone size at altitudes of interest.

ELFIN's second satellite compounds and expands its science. From close (<20min) along-track separations, the dual identical ELFIN satellites will determine, for the first time, the evolution of precipitating and trapped particles evolve as a function of L-shell on time-scales consistent with wave-particle interactions (minutes to tens of minutes). The agreement of observed temporal changes in the loss-cone fluxes as a function of energy and pitch-angle with modeled rates of EMIC wave scattering will be evaluated. The fill-rate results will then be compared directly with the measured precipitation rates of trapped radiation belt electrons, to determine quantitatively the global effect of EMIC waves on radiation belt losses. Thus, not only does the second ELFIN satellite increase the measurement cadence and number of storm event radiation belt crossings studied, but it also uniquely addresses the question of the global effects of EMIC wave scattering on radiation belt electron losses.

### 2.2 Secondary Science Objective

The secondary science objective of ELFIN (Table 1) is studied with only a modest increase in capability above the primary requirements (addition of the EPDI detector and extension of the electron energy range from 0.4 MeV down to 0.1 MeV). As this capability was available for the EPD heads from the Elfin-L program it was an easy decision to incorporate them also on ELFIN.

ELFIN's secondary science objective is to infer the location of field-aligned current (FAC) drivers relative to the physical boundaries of plasma populations during non-storm times, thereby advancing our understanding of processes related to the generation of those currents. These currents play a key role in magnetosphere-ionosphere coupling, establishing communication and energy transfer between these two regions.

Morphologically the dominant FAC systems are the region 1 (R1) and region 2 (R2) systems (Iijima and Potemra, 1976; Tsyganenko et al., 1993), each a few degrees in latitude, and together covering the entire auroral oval. Their sources are unknown, primarily because their generation region is spread over a large volume in Earth's magnetosphere, where local plasma gradients there are too weak and variable to be assessed using data from high altitude spacecraft. Region 1 generator candidates include (i) flow shear at the distant magnetopause/plasma sheet interface, (ii) the low latitude boundary layer, and (iii) vorticity or pressure gradients at the plasma sheet and its boundary layer beyond $15 R_E$. Although R2 currents are thought to be produced by azimuthal pressure gradients near the inner edge of the plasma sheet, their exact location is also unknown. Observations by low altitude spacecraft enable direct measurements of FACs and mapping them to magnetospheric populations using isotropy boundaries (IBs). IBs are the ionospheric latitudes at which a particle of a given energy and species will exhibit equal precipitating and trapped fluxes (Sergeev et al., 1982; Delcourt et al., 1996; Buchner and Zelenyi, 1989) due to equatorial pitch-angle scattering by high field-line curvature (Sergeev et al., 1993; Newell et al., 1998; Donovan et al., 2003). Because energetic (>50keV) particle precipitation is insensitive to auroral potential drops (<10keV), it is an excellent tool for remotely sensing magnetotail curvature based on this scattering phenomenon. The IB of >50 keV protons is near the strong (30-80nT) dipolar field at 5-8 $R_E$ (exact location depends on activity). Electrons >100 keV require smaller $B_z$ (~5nT) to be scattered. The ~1 MeV electron IB is near the periphery of the outer radiation belt (Imhof et al., 1977; 1997). Field-aligned current observations in the context of IB locations can thus be used to locate FAC generator regions in relation to such boundaries in the ionosphere and (through event-oriented mapping) in the magnetosphere (Kubyshkina et al., 1999).

Previous spacecraft (NOAA/POES, 1972-076B, and UARS) the data from which were used to describe the origin and morphology of IBs, were not equipped with a magnetometer to detect FACs. Spacecraft measuring the magnetic field (DE-2, FAST) had energetic particle-measuring capability that was either temporally limited or non-existent. As ELFIN is the first mission to make simultaneous magnetic field and pitch-angle resolved, high-cadence energetic particle measurements on the appropriate orbit to measure both FACs and IBs, its data can determine the location of FAC generators relative to particle boundaries such as the inner edge of the plasma sheet, the plasma sheet boundary, the low-latitude boundary layer, and the dipolar versus tail-like transition region.

At active (e.g., substorm) times, during both magnetospheric energy loading and substorm expansion/early recovery, the R1, R2 current systems move equatorward or poleward and undergo significant intensity changes. These currents are responsible for feeding magnetospheric energy to the ionosphere, powering its motional energy and Joule heating. Thus, it is important to understand their location and intensity as function of activity level and substorm phase. ELFIN's second satellite increases the cadence of these measurements to time scales consistent with substorm energy loading and unloading (i.e., less than 90min, the orbit period). Given a substorm recurrence time of 3 to 4hrs under moderate levels of solar wind input, ELFIN is expected to sample the R1 and R2 current systems and their mapping during hundreds of substorms during its minimum mission duration of 3 months. ELFIN will thus be able to study the spatial and temporal evolution of the R1 and R2 current systems and their source regions in the magnetosphere during substorm loading and unloading phases for hundreds of substorms, and during substorm expansion for dozens of substorms.

As mentioned in the introduction, during commissioning it was discovered that the EPDI detector on ELFIN suffers from a thick inactive area (akin to a dead layer), which prevents it from measuring <250keV protons. Therefore, as far as proton IBs are concerned, only the highest energies at the inner edge of the plasma sheet may be visible after pitch-angle (sector) and time (spin-period) averaging.

Neither type of averaging affects the ability to detect the IB, however. As the calibration of the ion detector is still under-way and data collections from the nightside have yet to occur, how well the secondary objectives can be addressed using energetic ions is still to be determined. As FGM and EPDE are operating nominally, FACs and electron IBs can still be used to address these objectives.

### 2.3 Synergistic Science

Several near-equatorial radiation belt particle and wave monitoring missions (THEMIS, Van Allen Probes, ERG, MMS, GOES) overlap ELFIN operations. Extension of the BARREL series of balloon X-ray measurements in the winter of 2019-2020, to measure precipitating electron fluxes over a wide range of longitudes and activity levels is expected to provide additional conjunction opportunities with ELFIN. Finally FIREBIRD II is still collecting data in orbit that are highly complementary to those of ELFIN. Although independent of these efforts, ELFIN is highly synergistic to them. Specifically:

- Together with the aforementioned equatorial missions, ELFIN will determine whether electron precipitation occurs at the same MLT and L-shell where EMIC waves are observed at the equator.
- Along with equatorial measurements of whistler mode waves, ELFIN will determine whether the observed loss-cone spectra – if characteristic of microbursts (Fig. 3) - are concurrent with equatorial chorus elements.
- While Van Allen Probes and ERG will be measuring the trapped populations' temporal variations and their *total loss rates*, ELFIN will be measuring *precipitation rates* allowing loss estimates from the *difference* between total and precipitation losses.
- In conjunction with those of equatorial satellites, ELFIN's EMIC wave measurements will reveal EMIC wave propagation properties, regardless of the wave effects on particle scattering.
- When combined with BARREL's inferred electron energy spectra over wide MLT ranges, ELFIN's actual measured electron energy spectra over a limited MLT range will provide a unique, inter-calibrated global precipitation dataset during storms.

These synergies will be achieved primarily as increased data volumes from ELFIN become available.

### 2.4 Primary Science Closure

ELFIN facilitates data distribution and analysis using a SPEDAS plug-in, and providing crib-sheets which that can be readily adjusted to the specific needs of each user. The fact that SPEDAS analysis tools are commonly used by many other missions (THEMIS, ERG, GOES, BARREL, etc.) enables full immersion of ELFIN in the H/GSO. On-line plots are already available at the mission web-site (http://elfin.igpp.ucla.edu under Science Overview → Summary Plots). Data dissemination through the Space Physics Data Facility will produce multiple data repositories for the mission. Data dissemination follows THEMIS's successful open-data policy.

Relativistic electron (>0.4MeV) scattering in the loss cone by chorus alone is quite benign, but when EMIC waves are added, precipitation is expected to become very significant (Fig. 6). Electrons at <0.4MeV cannot be in resonance with EMIC waves, which explains the deficient scattering in this range in diffusion models. Thus EMIC-only wave scattering can be clearly differentiated from chorus-only wave scattering, which does not possess a similar, sharp energy boundary at ~0.4MeV. When both wave types are operating simultaneously, appropriate modeling using observationally inferred amplitudes (from equatorial missions) is needed to deconvolve the spectra. We note, however, that the minimum EMIC resonance energy depends on observed EMIC wave spectral properties. It is also noteworthy that warm plasma effects caused by the presence of ring current ions can increase the minimum resonant energy of electrons due to EMIC waves to > ~1 MeV (Chen et al., 2011; Silin et al., 2011). By examining the temporal evolution of phase space density along the loss cone once per orbit (~90min or more often from dual satellite observations) ELFIN will determine electron ƒetimes over a wide energy range and make predictions about equatorial EMIC wave occurrence and wave spectral properties, to be checked by concurrent, equatorial missions.

Figure 7 shows modeled energy vs. pitch-angle distributions at a nominal, 400km altitude in the ionosphere for the cases of wave scattering discussed in Fig. 6. Although efficient scattering (filled loss cone) by EMIC waves is seen at >0.4MeV (the minimum resonant energy), it is absent when only chorus waves are present. ELFIN's electron energy versus pitch-angle spectra as a function of time relative to the start of storm time recovery will thus be compared with models to determine whether the electron minimum resonant energies and lifetimes are consistent with EMIC wave scattering. Precipitating spectra will be normalized to trapped spectra to ensure that temporal variations of the source do not affect the results. Trapped flux variations determine source variations with time, and indirectly also provide trapped flux evolution with time. ELFIN can thus determine EMIC wave-scattering efficacy and its importance for particle precipitation relative to other loss mechanisms.

### 2.5 Importance and Relevance to Heliophysics/Geospace Science

Quantitative tests of loss processes by the ELFIN team will enable thorough modeling of energetic electron losses by EMIC waves in the context of general transport models. Investigating such a potentially significant scattering mechanism, and advancing our understanding of electron losses in general, are just as important as understanding equatorial electron transport or acceleration because it is the combination of losses and sources that determines inner magnetosphere electron flux levels. ELFIN therefore fills an important gap in our understanding of particle and energy transport, as it enables (through inclusion into assimilation models, e.g., VERB) correct modeling of an important loss mechanism, namely precipitation. ELFIN is thus important for the NASA Heliophysics Division's and NSF's goals to understand fundamental microscopic physics as well as the global system behavior, as recommended by the 2012 Decadal Survey. ELFIN's approach to address electron losses system-wide by integrating its unique dataset and analysis with those of other NASA missions and NSF ground-based observatories addresses the primary recommendation of the Decadal Survey for system science with an integrated Heliophysics/Geospace system observatory.

## 3. Instrumentation

The two ELFIN instruments, the EPD and FGM, are packaged inside ELFIN's modular 3U CubeSat (Fig. 4 and Fig. 8), which adheres to CubeSat standards for P-POD deployment. ELFIN's key components are shown in Fig. 9 (integrated) and in Fig. 10 (before integration). The key components of ELFIN's interior are: (i) the instrument package except the FGM (which consist of the EPD sensor and electronics boards including the digital boards D1 and D2; Instrument Data Processing Unit, IDPU; and Switching Instrument Power Supply SIPS); they are located on the +X axis unit; (ii) the FGM electronics (FGE) board; the FGM boom (FGB); and FGM sensor which is mounted (stowed) inside the boom canister; they are located at the center unit; and (iii) the spacecraft avionics boards; batteries; radio; and antenna; they are located on the –X side unit. ELFIN is dynamically stable about the +Z axis after FGB deployment and it is power-positive under all attitudes.

The heart of the instrument electronics is the IDPU, an MSP430 processor running at 8.192MHz on an intellectual property (IP) core. Residing on an Actel FPGA (A3PE3000L-PQ208I by Microsemi), it also uses external non-volatile memory (two 256 kB FRAM chips for program storage and four 128 MB Flash memory chips for data storage). The processor runs an in-house, real-time operating system (RTOS) developed specifically for ELFIN. The $8.66 \times 9.32$ cm$^2$ IDPU board receives and stores instrument and housekeeping data, compresses data for subsequent downlink by the avionics boards, processes FGM spin-plane radial component data to generate a magnetic sectoring signal for the EPD instrument and packetizes the FGM and EPD data. It connects to other instrument boards (SIPS, EPD D1) by vertical mezzanine connectors and to the FGE by a twisted shielded pair through a 20 pin (M80-4662005) connector. Card-edge connectors run between the EPD D1, D2, preamplifier and bias boards, transmitting power and data. A coaxial cable and a flexible "flexi" cable communicate power and data to and from the EPD sensor, respectively. Instrument and mission requirements and specifications meeting those requirements are summarized in Table 3.

The only instrument with off-nominal performance is the EPDI. Designed to measure plasma sheet ions in the range 50-300keV at or near the inner edge of the plasma sheet, it has an off-nominal minimum energy of ~250keV because the Micron detectors received were laced with a 1.5 micron – thick layer of epoxy, used for wire bonding (the inherent 0.1 micron-thick window of the bare detector is therefore overwhelmed by the epoxy material atop it). This affects ELFIN's ability to measure plasma sheet ions (part of its secondary mission objective) but does not affect its ability to achieve its primary mission objective which is based on electron measurements. Investigations regarding the recovery of useful ion measurements at the inner edge of the plasma sheet are ongoing. The secondary objective can still be addressed using the FGM instrument and energetic electron measurements. Although the lack of ion measurements reduces the range of particle rigidities that can be used in IB boundary identification and magnetospheric model adaptation it does not prevent the mission from addressing and achieving closure of its secondary objective.

*Fluxgate magnetometer (FGM).* The FGM design was derived from the FAST (1996), ST-5 (2005), THEMIS (2007), MMS (2015), and DSX (2019) missions. The ELFIN sensors are identical to those on ST-5 (Fig. 11), where they had already flight-demonstrated the necessary dynamic range and sensitivity to EMIC waves (Engebretson et al., 2008). ELFIN's digital implementation of the fluxgate principle relied on a Sigma-Delta loop to perform sensing and feedback, by integrating the sensor as a component of the digitization cycle. First applied on the ST-5 magnetometer with commercial components, the design was flight-proven on THEMIS on an ASIC by TUBS and IWF, and was then implemented on DSX and MMS using rad-tolerant, low bit A/D converters and FPGAs. On ELFIN, discrete components were integrated on a multi-chip module (MCM) including the FPGA (A3PE600L). Calibration and electromagnetic compatibility (EMC) tests were performed at the magnetics and TVAC facilities at UCLA.

ELFIN's FGM deployment ~75cm away from bus sources of currents and variable magnetic sources occurs by means of a helically wound 2" BeCu strip, formed so that it is stored in a canister under tension. When released the strip unwinds into a slightly tapered, stiff, ~75cm-long, hollow tube (Fig. 11). The strain force releasing the potential energy in the wound-up, deformed coils in the canister causes this self-powered release. The design has been proven on numerous spinning missions, most recently on Cluster, STEREO (Ullrich et al., 2008) and THEMIS (Auslander et al., 2008). Two flight booms, procured from Kaleva Design Inc. (KDI) were operated and performed nominally on ELFIN. Although the boom has a resonance that exceeds three times the nominal (20 RPM) ELFIN spin frequency, it provides a comfortable 8% damping rate to damp nutation after boom deployment and reorientations within only minutes. A cord internal to the stacer provides an end-stop to the deployment before end of stroke and ensures knowledge of the deployed boom length. Release is initiated by actuation of a Ti-Ni shaped-memory alloy (SMA) manufactured by TiNi Aerospace. This "pinpuller" device operates by heating the alloy; hence, it provides a non-explosive actuation. Being retractable, the pinpuller allows testing in environments. The one-time in-flight boom release is actuated by the bus avionics unit by ground command. Deployment results in an FGM sensor orientation with its radially outward component nearly in the same orientation as during the stowed configuration (+Z in sensor coordinates, +Y in body coordinates, Fig. 9). As it is least-affected by the deployment this direction is used for EPD particle sectoring.

Two attitude determination and control system (ADCS) sensors and two torque coils (magnetorquers) are used to determine ELFIN's attitude and control its spin rate and spin-axis orientation. The ADCS sensors, two identical magnetoresistive magnetometers (MRMs; the second is used for redundancy), are integrated circuits (chips). One is mounted on the IDPU board in the instrument stack, and the other on the attitude control board (ACB) in the avionics stack. The two magnetorquers (rectangular spools made out of PEEK with yellow tape covering the wire in Fig. 10), one normal to the Y and one normal to the Z body axis, are used for ELFIN reorientations and spin-control, respectively. However, both the MRMs and the magnetorquers double up as a means of calibrating the FGM sensor in flight. During deployment, the stacer boom tip and the FGM sensor XY coordinates rotate relative to body coordinates, so it is important to devise a means of obtaining this rotation angle. This is done by comparing the FGM data with the MRM data. When MRM scales and offsets cannot be trusted (they change significantly on every MRM reset) calibration is aided by the generation of an artificial square-wave signal (of a fixed, commandable frequency in the approximate range of 0.5-2Hz) in the Y- or Z-normal torque coils. This creates a square-wave magnetic signal at the FGM on the order of ±100 nT strictly in the Y or Z direction, respectively (in body coordinates). Minimization of the orthogonal components in the FGM response provides the desired rotation angles.

*Energetic Particle Detector (EPD).* The EPD consists of two heads (Fig. 12): an electron head that measures 50keV to 5MeV electrons and an ion head that measures 50 to 5000 keV ions. A common set of front-end electronics implements processing similar to that done by the THEMIS SST instrument, but with mass and volume efficiencies realized from fast, low power (albeit low radiation tolerance) digital signal processing. Specifications can be found in Table 3.

*EPDE head.* The electron head consists of a stack comprising one 525μm front detector, two 1000μm, two 2000μm detectors, and an additional 525μm back detector, in that order. Each of the 2000μm detectors comprises two 1000μm detectors, biased separately but acting as a unit (their pulses are summed). These solid-state Si detectors form 6 channels (E1 – E6) measuring electron energies from 50keV to >5.8MeV (although it is an integral channel, last detector responds up to 7MeV based on GEANT4 modeling). A 10 μm Lexan foil is positioned in front of the stack. It is sandwiched between two 5000Å (0.5μm) Al layers. The foil completely blocks <500keV ions and light. The energies detected in each detector are summed using coincidence logic and are then binned in 16 energy channels with ΔE/E <40% (the highest >5.8MeV channel is integral). Front-end electronics and preamps, sensitive to radiation and electromagnetic interference are shielded by the detector structure or inside covers especially made for this purpose. Energetic particle measurements in the radiation belts are challenging because penetrating particles can contaminate the directional measurements. To suppress such noise (direct particle hits and secondary photon emission) EPD's detectors are in a vault with Ta (3mm) plus Al (9mm) walls. GEANT4 modeling of alternate multi-layer (including graded-Z) shield configurations (Pb, Ta, Sn or Al) has verified this design to be superior for both X-ray and secondary electron suppression, in agreement with past experience from OGO-5 design and data. Apertures (Fig. 13) include phosphor-bronze knife-edge collimators with a total thickness equivalent to that of the vault. These optics prevent one-bounce-scattered electrons from reaching the detector and suppress multi-bounce scattering. The coincidence logic, which also suppresses side-penetrating background, is most critical for the highest energies measured by the innermost detectors (E5, E6) where the counts are the lowest. Modeling of a realistic ion and electron environment shows a signal-to-noise ratio of 10:1 up to 5MeV, even without coincidence. Coincidence logic improves this ratio by a factor of 10.

*EPDI head.* This unit comprises two 525 μm Micron detectors (P1 and P2, same as for EPDE front and back detectors) and an electron broom magnet deflection system (0.28T gap field). The optics has a pinhole design to subject incoming particles to the large field at the gap. The broom magnet deflects electrons up to 500keV. Electrons >500keV deposit some energy into P1 but penetrate P1 completely and are thus also detected at P2. Using P2 in anticoincidence mode we thus also eliminate >500keV electrons from P1. Compensation magnets arranged in octapole fashion with the primary magnets result in fast fall-off of the offending field to a DC offset at the FGM sensor below 5nT. This is negligible compared to the natural variations of Earth's field. Low-noise electronics and the above technique permit ion flux measurements between 50keV and >5MeV, arranged at 16 energies with ΔE/E <40%. Although P1 is affected by sunlight, its fast electronics ensure quick recovery. Good isolation of the EPDI and EPDE signal chains prevents radiated emissions from the intense ion detector response to the sun from coupling into the electron side.

The data are binned by energy and sectored relative to the times of the ascending zero crossings of the derivative of the radial spin-plane component (along the +Y body axis, measured by the FGM sensor). These times correspond to the minimum value of that component (negative peak value of the magnetic field). Since the EPDE is mounted at a spin-phase 90º ahead of the FGM radial spin-plane component, these times also correspond to the EPDE head recording 90º particle pitch angles. The spin-phase sectored data are packetized for later compression (which occurs at times other than the data collection times to preserve computational resources). Ground processing with a-priori knowledge of the spin-axis attitude (from recent attitude data collection and processing) casts the EPD data in pitch-angle format using IGRF as the nominal field. Thus EPD processing does not rely on coordinate transformations or calibration of FGM data, allowing expeditious, automated creation of pitch-angle and energy spectrograms towards minimum mission closure. Such automated processing also forms the primary means of overview plot production that permits daily checkouts of the EPDE data by scientists, ensuring data quality and evaluating each downlink's importance for discovery.

## 4. Satellites

*Avionics.* The brains of ELFIN are on the ground. Bus and instruments are controlled by ground commands. Those are executed or distributed to the IDPU by a set of versatile, configurable avionics boards, located in the avionics stack on the –X side of the spacecraft (Fig. 8 and Fig. 14). Sensitive integrated circuits in the stack are partially radiation-protected by the four batteries. The avionics stack is thermally and electromagnetically shielded from the remainder of the bus. It comprises a flight computer printed circuit board (FCPCB), an attitude control board (ACB), two solar cell and battery printed circuit boards (SBPCB), and two interface boards dubbed "little et cetera" (LETC) boards: LETC1 and LETC2. The larger two boards seen in Fig. 14 are the "big et cetera" (BETC) board and the radio board, a He-82 model UHF transmitter/VHF receiver manufactured by AstroDev. The avionics stack is seen in the computer-aided design (CAD) drawing in Fig. 14, is also pictured (after application of electromagnetic interference shield and thermal treatments) in Fig. 10. The stack interfaces in the –X direction via coaxial cables through two quarter-wave sleeve baluns with the "-X panel" which hosts the antennas. The –X panel (also shown in the CAD in Fig. 8, and pictured in Fig. 10 but not in Fig. 14) hosts delay lines and the antenna deployment mechanism, or "tuna can" (thus named because of its shape). The fiberglass tape-spring antennas with beryllium copper inlay (the non-magnetic, conductive element) were manufactured by Loadpath. The deployment mechanism, which was built and qualified in-house, provides antenna restraint with pre-stretched spectra lines. These lines are wrapped over burn-resistors that actuate the release. The release is initiated by software timers after ELFIN's deployment from the launch vehicle. The FCPCB, ACB and SBPCB, hosting peripheral interface controller (PIC) microcontrollers, communicate with each other and with the remainder of the spacecraft; they were provided by the Aerospace Corp. and were programmed by UCLA students. The remaining non-commercial boards (LETC1/2, BETC, and –X panel) were designed and produced by UCLA staff and students.

The LETC1/2 and BETC boards provide power interfaces from the solar arrays to the SBPCBs, boost power conversion to support deployments and radio operations, interface between avionics and radio, and provide launch inhibits for communications and deployments. They also provide external access cable (EAC) interface to the spacecraft for testing; temperature sensor chains; battery heater logic; and interfaces between the avionics and two instrument boards: SIPS and IDPU.

The FCPCB is the central node of control and operations for the ELFIN spacecraft. It hosts two PICs: one for the flight computer and one for the watchdog. The flight computer processes ground commands, gathers and stores housekeeping data, constructs bus data frames, interfaces with the radio providing bus or IDPU data, and hosts the scheduler. It holds two copies of the software. The watchdog monitors the flight computer through "heartbeat" pulses exchanged between them and monitors critical telemetry; if the flight computer is stalled or critical telemetry limits are violated it resets the system. The SBPCB boards regulate bus power in response to ground commands. The ACB board collects telemetry from attitude control sensors and runs the control law algorithms (through the torquer coils). All PICs share the same communication protocol. The *Scheduler* receives stored scripts or uploaded command sequences that are authenticated using the SHA-1 hash, salt and key. It lives in flash memory and its schedule entries are time-stamped for execution; to save memory space some are reused by the spacecraft for repeat events. The FCPCB polls the Scheduler at every wakeup (once per 2.25 seconds). A *State Machine* on the ground, tracks the schedule entries and contents in the flight Scheduler and extrapolates its future state out to at least a week. Synchronization with the ground is performed frequently (at least daily) with all entry additions or modifications as part of the routine load files. The flight code is made tolerant to single event upsets by cyclic redundancy checks (CRC) and error correction. In-flight reprogrammability allows the software to be patched or rewritten and verified before the original flight code is to be replaced.

*ADCS.* The ADCS hardware comprises the two MRMs and the magnetorquers. Attitude determination begins by obtaining 5 to 8 snippets of 30 spins each, spread over an orbit, or a single continuous ~20min interval of MRM data during an orbit. Offsets, gains and spin-axis orientation are then determined by fitting these MRM data to the IGRF field. This results in the spin-axis knowledge to within <1°. This procedure is repeated once every two or three days. It is also repeated prior to and after reorientation maneuvers, to ensure successful maneuver execution, as the reorientation is typically as small as a few degrees. The attitude control operation must at least keep up with the precession of the orbit which (in a Sun-Earth-aligned system) rotates approximately 0.5°/day. Other torques (such as those caused by atmospheric drag or eddy currents) are less significant. Because the optimal science attitude calls for the spin-axis to be roughly along the orbit normal, the typical rotation of the spin axis by ~3.5°/week is accomplished by one such reorientation maneuver per week.

*Mechanical.* The ELFIN frame (light gray in Fig. 15) made from Al (6061-T6, anodized on the exterior and alodined in the interior) comprises four rails and two "top-hats" secured to each rail with two #6 fasteners. Originally based on Boeing's Tensor™ bus, a.k.a. the CS82 standard, the design was made available to ELFIN with permission, but was further mass-optimized for the specific ELFIN payload. Aggregate modifications over several years made the final design more similar to Cal Poly's ExoCube. This primary structure is the main interface with the Poly Picosatellite Orbital Deployer (PPOD). The PEEK torquer coil spools (dark gray in Fig. 15) do not contribute to the structural stability of the bus. However, the FGM stacer assembly (blue in Fig. 15), an integral part of the structure, provides shear strength in lieu of cross braces. Various assemblies and components attach to mounting positions on the rails and on the stacer assembly. The EPD assembly and instrument stack are mounted to the chassis via PEEK-to-Al brackets (seen in Fig. 9) that keep the assembly electrically isolated from the remainder of the bus at +4V, for efficient suppression of conducted noise.

*Power.* ELFIN's power system was derived from The Aerospace Corporation's AeroCube-4 hardware (Hinkley et al., 2009; 2010). The SBPCBs accommodate four lithium-ion 3.7V battery cells supplied and qualified by Aerospace (Halpine et al., 2009). These 2.4 Ahr Moli-Energy ICR-18650H cells are back-wired to minimize stray fields and are operated one battery string at a time. ELFIN's 20 UTJ solar cells manufactured by Spectrolab are also back-wired (using solar panel traces) and paired (opposite each other within each pair) to minimize current loops. There are four cells in the faces normal to the +Y and –Z directions and six cells in the faces normal to the –Y and +Z directions (Fig. 4, Fig. 16). Under a spin-axis orientation <15° to the Sun direction, boom shadowing and satellite spin significantly reduce operation of the +Y solar panels and minimize power input to the spacecraft. Under a nominal science orientation (spin-axis normal to the orbit plane), this situation occurs when the orbit plane is in the noon-midnight meridian. This orbit also entails the longest Earth shadows, further curtailing solar power input and increasing power utilization by survival heaters, exacerbating the low power availability condition for science. As the sought-after EMIC wave precipitation is minimized in such an orbit, the instruments can be kept OFF for approximately 1 month during each year (the orbit precesses approximately once per year), thereby conserving spacecraft power to be used towards battery heaters and housekeeping data downlinks.

A single, central spacecraft grounding point at the SBPCB minimizes current loops. Digital ground refers to that grounding point; the analog EPDE instrument ground level is at +4V relative to it. Each solar cell produces ~430 mA at 2.3 V, and pairs of solar cells power the battery cells by SBPCB control. Boost conversion produces the 5V bus that is then used to power the instrument stack, the avionics stack, the radio logic (after a 3.3V downconversion), and (after a second, 9V boost) the antenna cutters, the radio amplifier and the stacer deployment.

*Thermal.* The thermal design is passive, using judiciously selected thermal interfaces, coatings, tapes and Multi-layer Insulation (MLI) for component isolation and for reducing losses from openings and surfaces. Critical components are the batteries, which need to be prevented from discharging below $0^{\circ}$C, and the IDPU field programmable gate array, which should not exceed $40^{\circ}$C to prevent increased power draw. EPD detector operation above $30^{\circ}$C might also degrade instrument noise performance. The design provides for (non-magnetic) battery heaters to be operated in shadow as necessary to optimize the thermal operation based on in-flight performance data. Heater power allocation was accounted for in the power budget during the noon-midnight orbit; in-flight data are consistent with pre-flight model predictions of the need for low-level (duty cycled) heater power. The thermal design was implemented in thermal desktop using a trade-space thermal model informed by tests of surface treatment absorptivity/emissivity (a/e) and of thermal couplings. A simplified thermal balance test (STB) was used to validate our assumptions and to correlate the hardware components with an STB thermal desktop model. An additional benefit of this test is that it exercised the thermal vacuum (TVAC) facility and trained personnel in TVAC testing of the EM and FMs. The spacecraft remains power positive in all orientations. Even with a spin-axis-to-Sun angle of $90^{\circ}$ (worst-case cold conditions), there is operational flexibility to reorient the satellite away from orbit normal, thereby reducing potential heater power requirements and/or enabling science collection even for orbits near the noon-midnight meridian. Similarly, operational workarounds can reduce in-flight temperatures under worst-case "hot" conditions (spin-axis-to-Sun angle ~$69^{\circ}$, occurring near a dawn-dusk orbit plane) should in-flight performance data indicate such a need. So far, in-flight component temperatures have been nominal and no such workarounds have been necessary.

*Resource budgets.* ELFIN's mass and power budgets are shown in Tables 4 and 5 respectively. The data, which are based on measurements taken during EM or FM testing, are based on the pre-ship review. In-flight power consumption and generation data are consistent with these measurements.

## 5. Ground Systems

ELFIN communicates with the ground via its UHF (downlink, 437 MHz, commandable 9.6kbps or 19.2kbps rates) and VHF (uplink, 146 MHz, 9.6 kbps uplink rate) Astrodev He-82 radio, operated under a Federal Communications Commission amateur license. ELFIN-A and –B have slightly different downlink frequencies, such that they can be separated when both are above the ground station, and a fairly isotropic gain such that they can communicate in any orientation. The link budget is shown in Table 6. The ground segment consists of the Mission Operations Center (MOC), Earth Station (ESN), Science Operations Center (SOC) and Flight Dynamics Center (FDC), all at UCLA. The primary ELFIN ground station (the only one used for uplinks), is at UCLA. Located on the roof of Knudsen Hall (call sign: W6YRA), it has two uplink antennas (a primary, Knudsen South, and a backup, Knudsen North, each consisting of 2 VHF yagis) and a single downlink antenna (consisting of 4 UHF yagis on the same Knudsen North pedestal as the 2 VHF yagis), as depicted in Fig. 17. Additional downlinks of scheduled passes are provided by NASA/Wallops (NASA's Wallops Flight Facility) and the Stellar Station (Japan), both using multi-meter diameter dish antennas that provide long-duration downlinks with high throughput. Unscheduled ELFIN beacon relay support is provided by SatNOGS a global network of satellite ground stations, operated as an open source participatory project (since 2014, SatNOGS has surpassed GENSO in international participation, global coverage, and capability).

The MOC performs planning and scheduling of passes, command generation and uploading, telemetry monitoring, analysis and trending. A multi-mission, multi-spacecraft center, it interfaces with the UCLA ESN and other ground stations, sends commands for uploads, requests passes, and receives and processes data. The SOC obtains level-0 (L0) data from the MOC and casts them into L1 files in common data format (CDF). These files are then distributed to the community for analysis. Once calibration and extensive analysis of all instruments has been accomplished, likely near the end of the mission, L2 files are slated for production. The SOC also produces ephemeris (position, attitude, spin rate) and calibration files as well as overview plots that serve the science community. The FDC receives one or more deliveries of two-line element files (TLEs) from the Joint Space Operations Command (JSpOC) daily for each spacecraft. It uses those files to produce definitive daily orbit information and it periodically produces short term and long term orbit projections. It also processes MRM data and provides attitude information based on analysis of that data. The FDC conducts reorientation and spin-up/down maneuver planning using a maneuver simulator (an orbit and attitude integrator that uses Matlab's Simulink and an internal representation of Earth's IGRF model) based on spacecraft current attitude, target attitude, torque coil performance and orbit information. Each maneuver is broken-down into individual segments placed judiciously over one or more orbits. The simulator determines the optimal location for performing each maneuver segment to take advantage of the strongest Earth field available in the desired orientation, thus minimizing total energy consumption for a given total reorientation. Typically ~10 individual reorientation maneuvers of 5min each, spread over three to four orbits are needed to precess the vehicle by 3.5º once per week. Only one spin-up maneuver per week is needed to maintain spin-rate near the desired 20RPM. The FDC also performs trending of drag coefficients that are then used to predict orbit projections and orbit lifetime. Predicted orbits are also routinely and periodically released to the science community to aid future conjunction planning.

The ESN executes passes according to MOC requests, transmits commands and receives raw telemetry from the spacecraft, which is demodulated, decoded and sent to the MOC (as L0 raw data files). The ESN is highly automated, with a graphical user interface (GUI) controlling the station remotely to track each satellite and log local station engineering data, including status of the radio, amplifier, remote camera, and temperatures. It operates using a Ground Support Electronics Daemon (GSED) server, the central relay point between the MOC and the ESN. The GSED, which permeates real-time command generation, the graphical planning console and the ESN daemon is part of the ESN network. As asynchronous Python server, it was built using the Curio asynchronous library. It dispatches commands to the spacecraft, manages command load files, authentication, security, requeueing and aborts. The ESN daemon is then responsible for commanding the ground station to execute operations such as tracking and up- or downlinks. The commands and load files are issued by the MOC commander and are then handled by the GSED and passed through the ESN interface manager to the ESN daemon. Played back by the station in their most primitive form, the downlinked files are raw signal (waveform) files stored as In-phase and Quadrature data (IQ.wav or IQ.dat files) and can be passed straight to the MOC. However, the nominal situation (for the UCLA ESN and Wallops stations) is for the files to be demodulated using software-defined radio applications on the ESN side, and to output the result to a port which the GSED can understand as a byte sequence and satellite id. The exception is that Stellar Station transmits IQ files that are demodulated at the MOC manually. Automation is expected to take over this operation soon. Nevertheless, GSED, handles both cases well, and is being used by all ELFIN ground stations that provide scheduled pass support.

## 6. Mission and Science Operations

*Mission Operations.* ELFIN's Mission and Science Operations are collocated at UCLA (Fig. 18). The MOC performs pass planning and activity scheduling, command generation, frame parsing, limit-checking, telemetry displaying (near-real time), analysis and trending. It comprises several workstations loaded with a suite of software. Operations rely on orbit predicts and command sequences for long-term science collection and contacts. Interleaved with these are near-real-time commands, as needed to obtain spacecraft health and status or play back specific data under off-nominal conditions. Orbit propagation is done using the Satellite Took Kit (STK), with a Python interface to manage, use and update the TLEs, and populate events tables for communications, next science zone (for data collection), Sun and shadow intervals, etc. Interactive Data Language (IDL)-based attitude determination code is spawned from within Python. It analyzes MRM data collections and returns to the mission operations the definitive attitude information for ingestion into state files as well as for attitude propagation within maneuver planning tools. The remaining packages are home-grown and written in Python.

The Planner is used to schedule operations activities for each spacecraft. It plans those activities long-term (at least one week out), and modifies short-term (daily) schedule entries subject to post-maneuver conditions, or science analysis, or recent pass performance. It is synchronized with the flight computer Scheduler. Taking the high-level (abstract) operations schedule and orbit events in the database the Planner creates a detailed schedule of activities for spacecraft and instrument operations and housekeeping passes, holds and updates events trigger tables, and notifies the user of operational constraints. It also accesses and visualizes spacecraft resources (power status, temperatures, ADCS sensor data, memory) and downlink opportunities.

The Translator translates scheduled operations activities in the Planner to spacecraft commands, and manages on-board scripts and the execution schedule.

The Commander generates command load files from the translator output. It also performs real-time commanding of the spacecraft via a command line and button interface, as needed during passes. Finally, it controls GSED configurations and settings on the ESN side.

The State Machine works hand-in-hand with the remainder of the software to help visualize the past, current and future state of the spacecraft. Using a GUI, it displays the latest real-time data it receives, and graphs live telemetry trends as the data come down, or as projected in the future by the Planner. It determines the spacecraft's state from incoming packets, stores state information in the database, tracks memory usage and state, and is optimizes memory reusability.

*Launch and Early Orbit Operations.* Upon ELFIN's injection into orbit, the separation switches turned ON the spacecraft, and ELFIN entered "early orbit operations" mode, an autonomous mode that prevents transmission and actuation of deployables for the first 45 minutes in orbit and initializes the flight computer. The FCPCB checked the non-volatile memory and real-time calendar clock to determine status, enabled and executed antenna deployment, and then enabled beaconing every 90 seconds. Signal acquisition and command uplink for scheduled passes to downlink health and safety information followed. After spacecraft checkout, de-tumble and initial spin-up to ~10 RPM were performed. These was followed by magnetometer boom deployment on ELFIN-A; the boom remained stowed for a two additional weeks on ELFIN-B. This established a 0.06 m/s differential velocity between the two probes that provided an along-track interspacecraft separation (ELFIN-A leading –B) time of ~15 min per year, during the first year of operations. In the future differential drag operations may still be implemented by changing spacecraft attitude to further modify the inter-spacecraft separation. Following separation velocity increase, reorientation to science attitude and final spin-up to science spin rate took place. As per requirements, the spin plane contains Earth's field during at least one (or more, depending on inclination) auroral crossing per orbit. All attitude operations took place using the MRMs. Instrument commissioning was followed by initial instrument data transmittion, status evaluation, and software modifications to spacecraft and instrument operations to ensure optimal operations. Spin-ups and reorientations were done to increase proficiency with the algorithms and operations.

Even though the on-orbit operation of ELFIN has been successful, several issues have caused delays in routine science implementation. The most important has been with implementation of the ADCS software that predicts reorientations, a Matlab Simulink code that propagates attitude information subject to torques applied by torque coil action at a given orbit location. In mid-2019 it was discovered that the simulated and executed reorientations were not performing to the degree of accuracy desired. It was later discovered that this was partly due to MRM offsets and gains that were unstable (hence inserting errors in the initial attitude) and partly due to attitude propagation errors within the propagator. These issues were resolved in late 2019: MRM offsets and gains were stabilized by preventing reset of the MRM sensors, and attitude propagation in the Matlab code was identified and fixed. The end-to-end reorientation process is currently being tested in orbit.

*Nominal operations.* Routine operations entail generating weekly spacecraft and payload activity plans after checking constraints and resources (e.g., power and ground station availability). They also entail building and checking command loads; monitoring spacecraft health and safety; data trending, clock synchronization and downlinks of science and engineering data; planning and performing weekly spin-up and reorientation maneuvers; and evaluating the results to ensure successful execution.

*Contingency operations.* The MOC monitors every pass and alerts the flight operations and flight dynamics teams of limit violations (alerts persist until acknowledged). An on-call controller evaluates the severity of the anomaly and notifies all flight operations and other members (software, bus, and instrument engineering). If the anomaly is sufficiently severe, the Mission Operations Manager and the Principal Investigator are notified. A tiger team made up of the cognizant engineers, flight operations, and flight dynamics teams (as needed) is assembled for critical contingencies. A report is written within hours of the event and is updated to reflect minutes and actions from tiger team meetings. If the consequences are of at least moderate likelihood and mission-critical, the PI informs NASA and a summary of actions taken is provided. If likelihood and severity are deemed high or if resources within the team are insufficient to resolve the issue, call for support from NASA and other teams is a required part of the action plan.

*Science Operations.* Primary science closure is based on EPDE pitch-angle and energy-resolved spectrograms of radiation belt crossings (each dubbed a "science zone") at a cadence of at least once per four orbits (faster than the timescale of storm-time radiation belt flux evolution). There is sufficient power in dawn-dusk orbits (when, for nominal, orbit-normal spin-axis orientation the spin-axis-to-Sun angle β is >30°) to collect 16 times the minimum data volume (Table 5). Every science zone (intended to cover the radiation belts, L-shells between 2.5 to 7.0), lasts 5min and is captured with full EPDE spectrograms (16 sectors per spin, 16 energies per species) and FGM collection at 10 samples/s. FGM collection starts 2min prior to the science zone to ensure that the FGE electronics is thermally stable. Compression, which accompanies all transmissions (nominally), is commanded to place prior to planned data downlink to only the data to be transmitted (saving on board computations and battery power). Operational flexibility exists in transmitting survey data that contain field-aligned and perpendicular spectra in every science zone. Survey data constitute a different form of (lossy) compression, while the full data remain intact on-board for future downloads if deemed interesting by the scientists. Two science modes are thus envisioned: Slow (or Survey) mode entails transmission of EPD Survey data and Fast mode entails full pitch-angle and energy EPD data. FGM collection at 10 samples/s accompanies both modes. Although there is sufficient power and on-board memory to collect, compress, and store the resultant data volume (Table 7), transmission to the ground is the limiting resource for ELFIN data acquisition. Two passes per day per spacecraft (one at the UCLA station and one at Wallops Island) allows transmission of the maximum data (~1MB per day per spacecraft) to be transmitted, based on 60% downlink efficiency, at a 19.2k rate. Stellar Station provides additional coverage for potential down-times of either of the two nominal stations; UCLA alone can downlink about half the maximum mission data, well above the minimum requirements. Mission operations personnel carefully watch space weather alerts and indices (such as the Kp and AE index) to determine whether a magnetic storm is impending and (subject to confirmation by the scientist in charge, the "Tohban") when to optimize collections should schedule conflicts require non-optimal station availability.

Routine science operations entail planning and executing the above science data collection plan and processing the raw packet data using Python scripts. The raw data packets are time ordered, overlap-deleted, and cast into L0 packet files within 4 hours after receipt. The SOC routinely processes the L0 files into L1 daily files that are in common data format (CDF) and still contain raw (uncalibrated) quantities. The L1 files, the primary product intended for science data analysis in conjunction with the calibration files are produced automatically once new files are downlinked. The L1 files are quite stable but the calibration files are updated when scientific analysis produces better-informed calibration products. The calibration files, separate files in text format, are thus updated less frequently. The updates are transparent to the analysis code; the corresponding science software is also routinely updated as necessary to reflect potential changes to the calibration procedures (note that simple updates of calibration parameters do not necessitate code updates). L2 files are expected to be generated closer to the end of the mission, when calibrations are stable and calibrated quantities in various coordinate systems and physical units are to be produced and disseminated. All L1 files and software are expected to be released to SPDF once routine operations commence in mid-February 2020. L2 files will be released upon their generation and checkout near the end of mission.

The SOC also automatically produces overview plots for first-look and subsequent data validation by the Tohban. The orbit plots consist of definitive and projected orbit information in geographic and SM coordinates (Fig. 19). Automated processing of L1 instrument data files (using preliminary calibration files, hereafter referred to as "preliminary") results in instrument overview plots. EPDE plots are already in production (Fig. 20). Additional panels containing FGM and EPDI plots will be added in the future. Because automated plot production occurs within minutes of data download, it allows for a quick status check of instrument health by an appointed scientist-in-charge, otherwise known as "tohban". The biweekly tohban service by members of the science community serves the dual purpose of intimately familiarizing team members with the data and of recognizing scientifically interesting events that merit further study.

*Mission Status and Plans.* The EPDE and FGM instruments on both satellites have been commissioned. Operations are being streamlined for routine data collection starting in February 2020, once the orbit plane moves out of the noon-midnight meridian. Data collections are expected to cover the remainder year of orbit life (reentry is expected in early Spring 2021), and to provide complete coverage (24 hrs) of magnetic local time. Analysis software that provides high quality pitch-angle and energy spectrograms and removes the Earth's intense field to reveal low frequency (EMIC) waves is under construction and will be released soon (overview plots provide preliminary information). Data collections from three prior storms have already provided sufficient information to enable closure of the prime mission objective. Examples of data collected are presented in the next section. The EPDI instrument on both satellites operates nominally except that the solid state detectors suffer from a large dead layer that prohibits detection of ions below 250keV. Higher energy ions, up to multi-MeV, are being collected at lower latitudes, albeit with a geometric factor that is small for the lower fluxes expected at such high energies. Thanks to the low background of the detector, summing over energies and pitch-angles can still result in useful information. Those modifications are still to be performed in orbit; when that occurs, EPDI collections will also be streamlined during the last year of ELFIN operations.

## 7. Data Analysis

ELFIN's analysis, done in IDL using heritage routines from THEMIS, is a SPEDAS plug in (Angelopoulos et al., 2019). The software benefits from coordinate transformation and Tsyganenko magnetic field models that include IGRF, field line mapping (including ionospheric or equatorial projections), and access to ancillary data from numerous space and ground observatories. Because THEMIS, ERG, MMS, and many ground-based observatories are accessible by the same software, this approach provides a unified way of analyzing ELFIN data in the context of the H/GSO.

Figure 21 shows a plot created by these routines for a subset of the data shown in Fig. 20. It is intended to reveal not only several interesting features of the dataset that will be further studied and presented in future publications, but also to demonstrate some remaining issues with EPDE calibration that still require attention (we anticipate that they will be corrected in the future). These issues, however, will not prevent the mission from achieving its prime science objective. The figure shows energy spectrograms of trapped (Panel A) and precipitating (Panel B) fluxes, the ratios of parallel (down-going) to trapped fluxes and anti-parallel (upward-moving) to trapped fluxes (Panels C and D, respectively). It also shows pitch-angle versus energy spectrograms from energy bins in the ranges of 50 to 160keV and 345 to 900keV (Panels E and F) and the magnetic field from IGRF at the bottom panel in the same coordinates as in Fig. 20 (horizontal magnetic North, vertical Down, and horizontal azimuthal magnetic East) plus the total field magnitude, T. Note that the IGRF was used to compute pitch-angles in these plots, based on satellite location and attitude. However, the on-board magnetic field measured by the FGM is used by EPD for the purpose of sorting, or binning, the particles in sectors in relation to the locally measured magnetic field phase angle. Although there are 16 sectors per spin, the number of sectors is commandable and can be increased to 32. There are 16 energies per sector, which can also be changed by command. The MRM cannot be used for such sorting because of its high noise level, whereas the ability to bin the data is an effective means of data compression, important for optimizing the science.

Panel A in Fig. 21 shows significant trapped fluxes up to 2MeV. Intensifications of the trapped flux were observed at several times, in particular at ~1620:45, 1621:37, 1621:48, and 1621:54UT. Each intensification lasted for at least one spectrum and approximately two to three consecutive energy spectra. The flux variability during consecutive spectra was high which suggests temporal aliasing of the flux level. Since the flux ride-and-fall time-scale is ~1.5s or less, these are likely microbursts. The parallel (downward) fluxes in Panel B exhibit similar behavior as the trapped fluxes except their flux is typically lower. The precipitation is clearly recognizable also in the pitch-angle spectra over the select energies depicted in Panels E and F. This means that precipitation occurs over a broad range of energies, at least up to 1MeV throughout this time, and up to 2MeV during the microbursts.

To better quantify the precipitation and contrast it with backscattering, we examine the ratio of parallel (precipitating) to perpendicular flux in Panel C and the ratio of antiparallel (backscattered) to perpendicular flux in Panel D. It is evident that both precipitation and backscattering occur at a low but non-negligible level (on the order of ~10% of the trapped flux in this case) at all energies and times, but that both precipitation and backscattering intensified surrounding the times of the microbursts (and can even reach levels close to those from trapped fluxes). As the data collection took place at low latitudes (L-shells around 3-5) no field-line curvature effects, only wave scattering, could be involved in this scattering process. Most times the precipitating to trapped flux ratio in Panel C is decreasing as function of energy. This suggests that either the resonant interactions dominate at low energies, or that the equatorial pitch-angle distribution around the loss cone has a sharper gradient at larger energies. Thus chorus or kinetic Alfvén waves which can scatter >100 keV electrons through the drift-bounce resonance (Chaston et al., 2018) are likely implicated in such scattering. However, there are two times when this ratio increases with energy. These times are around 1621:37 and 1621:48 when that ratio at 0.9 – 1.5MeV is around 0.8-1.0, and much greater than at 0.35-0.6MeV where that ratio is around 0.4-0.6. In particular in the case around 1621:48 UT this situation persists for 7 consecutive energy spectra, or 2.5 spins, which suggests that this is not due to temporal aliasing but a persistent feature of the data. Such energetic electron distribution is likely due to EMIC waves. Determination of the resonant energy of such waves observed simultaneously on ELFIN or at conjunctive observations by other missions at the equator should be able to further confirm or reject this assertion, and is part of the primary objective of the ELFIN mission.

The backscattered-to-trapped flux ratios are also significant and comparable to the precipitating-to-trapped flux ratios. For example, near the times of the 1621:37 UT and 1621:48 UT microbursts, the two ratios differ by only a factor of 2 to 4 depending on time and energy (i.e., backscattered fluxes can be 25-50% of precipitating fluxes). This effect is also clearly visible in the pitch-angle spectra plots (Panels E and F). Surprising given results from atmospheric interaction models of such backscattering (Marshall and Bortnik, 2018), this suggests either that such backscattering is far more efficient than previously thought or that other processes, such as local acceleration by low-altitude waves may be responsible for the observed flux increases. These ELFIN observations need to be further quantified and compared with models.

It is interesting that at ~1621:54UT to observe that the parallel-to-trapped flux ratio exceeds unity for several spectra around the last microburst in the sequence over a broad range of energies around 0.2-1MeV. This is corroborated by the pitch-angle spectra at around those times. Because the microburst trapped and precipitating flux can be quite variable over a half-spin (1.5s) time-scale, temporal aliasing cannot be immediately dismissed as the origin of this variability. Further analysis of the timing of individual sectors responsible for the trapped flux measurements is necessary to further investigate the origin of this effect. If validated by further analysis, this observation would also suggest that local acceleration of particles could be responsible for the highly field-aligned nature of the precipitation, because otherwise a near-equatorial source would experience angular broadening as it propagates along its propagation to the ionosphere due to the magnetic field strength increase.

The very high data quality from the primary ELFIN instrument enables studies that can result in closure of the primary mission objective. Because no previous mission has had ELFIN's simultaneous high energy, pitch-angle and time resolution, the dataset collected also opens new questions not envisioned when ELFIN was proposal. As the second ELFIN satellite, separated by ~15min (presently), bears data of similar quality, inter-spacecraft comparisons at such short timescales provide a unique dataset that can resolve the temporal variability (rate of change, spatial evolution) of the phenomena discussed. The anticipated year-long routine data collections from ELFIN promise to go far beyond ELFIN's original goal, to understand the role of EMIC waves in relativistic electron scattering.

In the analysis, however, care must be taken to overcome several pitfalls associated with the data collections that are still under investigation. *First*, some "pixelation" evident in the pitch-angle spectra (Panels E and F) bespeaks an offset in the spin-phase sectoring of the original data on the spacecraft. This results in similar pixelation in the energy spectra: the 0-180º pitch-angles have a flux at fixed energy that is slightly different from that of the 180-360º pitch-angles, even at times when the temporal variability of the perpendicular flux (Panel A) is small. This is not caused by temporal aliasing, but by a timing uncertainty in the FGM zero crossings relative to the sector start times. Once on board data collection within a science zone commences, the timing stabilizes to a fixed offset, but the offset can vary from one science zone collection to another. This timing offset varies with spin rate by as much as 0.4 seconds (as much as two sectors worth) but for a fixed spin rate it varies by ±0.038 sec (by as much as ±4.5º). Though feasible, a software patch to determine the actual FGM timing entails significant programming and testing effort. A simpler approach to minimize this effect in orbit was to stabilize the spin rate (performing routine spin-ups once per week), measure the typical offset, and insert an opposite offset to the sectoring software in the IDPU to balance the systematic offset and ensure that the B-field lies in the middle of the first sector in the spin. This is expected to remove the systematic part of the timing offset. The data from past collections and future ones are processed, and the offset is determined by matching pitch-angle spectra on the two sides of the magnetic field. This offset then becomes part of the data calibration that produces on-the-fly calibrated data from L1 data. Referred to as "regularization", this process maps the data in each spin to a regularized pitch-angle grid with perpendicular fluxes centered at 90º and parallel ones centered at 0º and 180º. The offset used in the processing in units of whole sectors (each sector being 22.5º) plus additional degrees is displayed in the overview plots. For example, in Fig. 20 the offset detected and applied was 2 sectors plus 1.9 º, or a total of 46.9º. This offset value was also used when processing the data for Fig. 21. The in-flight offset implementation is currently under testing, in-flight. If it is successful in producing high-quality spectra (i.e., reducing the pixelation to negligible levels), a further patch of the sectoring process could be avoided. Nevertheless, even with remnant pixelation, the data quality is sufficiently good, as discussed earlier, to satisfy the minimum mission objective.

*Second*, the FGM instrument performed nominally in flight, but its offsets, gains, and rotation matrix from instrument coordinates to body coordinates require further calibration. The rotation matrix is needed to determine FGM offsets and gains. This matrix is unstable, however, due to the boom's torsional oscillations, which occur because the Sun illuminates the boom from a different direction upon each rotation (the thermal response of the boom is faster than the spin period because the stacer boom is not blanketed). These oscillations are thus driven at the true (sidereal) spin period of the satellite. The measured magnetic field is also used to determine the apparent spin period from zero crossings of the radially outward spin-plane component (since the field changes along the orbit, this is a synodic period). This happens using the FGM sensor on board for subsequent EPD particle sectoring and using the MRM sensors on the ground for controlling the satellite spin period to within limits established for optimal

EPD sectoring. The two periods, sidereal and synodic, both around 0.33Hz, are slightly off of each other creating spin-tone harmonics that interfere with the ability to measure waves in the 0.5 to 2 Hz frequency band, which is associated with EMIC waves. Determination of the rotation matrix, however, is aided by an artificial square-wave signal injected in the Y- and Z-normal torque coils (separately) that enables us to determine the orientation of the FGM sensor in body coordinates. These tests have been successfully performed and analysis of the data is ongoing. Once rotation of the FGM in body coordinates has been performed, gains and offsets are expected to be determined, which should result in ambient magnetic field removal and EMIC wave identification. The FGM calibration for wave identification does not impact the ability of the mission to achieve its primary science target. FGM data collections during EPDE data acquisitions and downlinks continue, and the FGM data will be released once calibration routines become available.

*Third*, the EPDI instrument's in-flight performance is still under investigation. Tests of the EM model using alpha sources at two different energies have revealed that the dead layer of the detectors, around 1.5 microns, corresponds to an energy loss of ~200keV, rendering the minimum detection energy ~250keV (at least for the EM detector). For this reason, new in-flight operations in the inner belt are being considered, to determine the efficacy of the instrument at higher energy ranges, in particular after summing over energy and pitch-angle to enable sufficient counts over time. Once this has been confirmed, routine collections over intervals when fluxes are expected to be significant will be inserted in the ELFIN operations plan. Although the EPDI operation at higher than originally planned energies does not affect the primary mission objective, it may result in new science quite different from what was anticipated in the ELFIN proposal as part of its secondary objective.

In summary, the overview plots from ELFIN already demonstrate a mission with high potential. The analysis code is being exercised and readied for release pending final calibrations of the EPDE instrument. The calibrations entail in-flight testing that can reduce or completely remove pixelation in the spectra arising from the timing uncertainty in binning the data in flight. The resulting delay is warranted given that the routine data collections are to commence in February 2020 (presently ELFIN is in a noon-midnight meridian orbit, which is less optimal for power generation as well as EMIC-induced scattering). Routine data collections are expected to last for a full year covering the full range of MLTs and providing a very rewarding dataset.

## 8. Concluding Remarks

In the last ten years, rapid development in commercially available CubeSat hardware has increased the scientific potential of missions involving single or multiple CubeSats. Additionally, low-cost launch opportunities, underwritten by sponsors, such as NASA, have ushered in a new era of space exploration by capable, rapid-turnaround missions with little-to-no overhead from agency reliability requirements. This represents a unique opportunity for addressing focused scientific objectives to fill knowledge gaps or support larger programs, but also executing high-risk, high-payoff experimental space research. Finally, it represents a novel way of supplementing the educational experiences of (primarily but not only) science, technology, engineering and math (STEM) students: as on-line learning replaces traditional teaching models, higher-education institutions will have to reinvent themselves as places that can support advanced, informal educational experiences. CubeSats motivate students to excel in real-world, challenging projects, in laboratory and peer-group settings, but in a supportive environment. Such training is highly-valued in subsequent industry or research employment. Finally, CubeSats enable low-barrier to entry of academia into space research, and widen the opportunity to cultivate and tap talent in diverse geographic and sociocultural backgrounds.

ELFIN was designed, structured and executed with full appreciation of the aforementioned emergent transformation in space research and education. As a case-in-point, ELFIN's primary science objective, is a key question in space physics. Resolving whether the mechanism responsible for relativistic electron precipitation is EMIC waves and whether those waves result in (spatially) structured or (temporally) transient depletion of the outer radiation belt cannot be answered without multiple, low-altitude, pitch-angle resolving satellites. ELFIN's two spinning satellites are the first to have the pitch-angle and energy resolution required to address this objective, and contribute to our understanding and prediction of space weather. In conjunction with Van Allen Probes, THEMIS and ERG, the ELFIN satellites can address additional questions, including the energy range and spatial distribution of chorus-driven precipitation, and the net precipitation and backscattered rates in the range 50keV to several MeV. The mission was designed, developed and is now operated, by a total of >250 undergraduates in mostly (though not-only) STEM disciplines and from diverse socioeconomic backgrounds. Most have since already placed successfully in industry, in research and development organizations, or in academia. Their experiences have provided a track-record of success for future flight hardware evolution, and represent a model for informal, inter-disciplinary, higher education. Finally, ELFIN is reaffirming the high-return science and operations which can be obtained from multi-satellite CubeSat missions, and brings us closer to the realization of multi-satellite constellations, an important means of exploring the Heliosphere.

**Acknowledgements.** We acknowledge support by NASA award NNX14AN68G (5/22/2014 – present) and NSF award # 1242918 (9/24/2014-7/31/2019). We are grateful to NASA's CubeSat Launch Initiative program for successfully launching the ELFIN satellites in the desired orbits under ELaNa XVIII. We thank the AFOSR for early support of the ELFIN program under its University Nanosat Program, UNP-8 project, contract # FA9453-12-D-0285 (02/01/2013-01/31/2015). We also thank the California Space Grant program for student support during the project's inception (2010-2014). We acknowledge the critical contributions and talent of numerous volunteer team members (more than 250 students contributed to this program since its inception) who made this challenging program both successful and fun. The editorial assistance of Judy Hohl is greatly appreciated. Last but not least, ELFIN would not have been possible without the advice and generous time contributions from a large number of the science and technology community who served as reviewers, became unofficial mentors, or were the bouncing board of ideas during trade studies or tiger team deliberations. Special thanks go to UCLA's: Steve Joy and Joe Mafi; The Aerospace Corporation's: Eddson Alcid, David Arndt, Jim Clemmons (now at UNH), Chris Coffman, Joe Fennel, Michael Forney, Jerry Fuller, Brian Hardy, Petras Karuza, Christopher Konichi, Justin Lee, Pierce Martin, Leslie Peterson, David Ping, Dan Rumsey and Darren Rowen; NASA GSFC's: Gary Crum, Thomas Johnson, Nick Paschalidis, David L. Pierce, Luis Santos and Rick Schnurr; NASA JPL's: Matt Bennett, Olivia Dawson, Nick Emis, Justin Foley, Travis Imken, Allen Kummer, Andrew Lamborn, Marc Lane, Neil Murphy, Keith Novac, David R. Pierce, Sara Spangelo, and Scott Trip; AFRL's: Travis Willett, David Voss and Kyle Kemble; UCB's: Peter Berg, Manfred Bester, Dan Cosgrove, Greg Dalton, David Glaser, Jim Lewis, Michael Ludlam, Chris Smith, Rick Sterling, and Ellen Taylor; Tyvak's: Bill McCown, Robert Erwin, Nathan Fite, Ehson Mosleh, Anthony Ortega, Jacob Portucalian, and Austin Williams; Cal State University Northridge's: James Flynn and Sharlene Katz; JHU/APL's: Jeff Asher and Edward Russell; Montana State University's: Adam Gunderson; Space-X's: Michael Sholl; Planet Labs': Bryan Klofas; and LoadPath's: Jerry Footdale.

| | Science Objective | Measurement Strategy | Instrument Requirements | Mission Requirements |
|---|---|---|---|---|
| **Primary** | Determine whether EMIC scattering is the primary loss mechanism of radiation belt "killer electrons", or if other mechanisms are also important. (Minimum). Study at least 2 storms; during the course of one storm. 2. Measure energy and pitch-angle spectrum with observed low altitude waves to ascertain presence of EMIC waves (Baseline). Is global loss rate due to EMIC? (STAR) | 1. Measure relativistic electron loss rate from consecutive passes of radiation belts, >60° magnetic latitude, once/90min. 2. Measure energy and pitch-angle spectrum of precipitating electrons to determine whether it bares typical signatures of equatorial scattering by EMIC waves, or other processes. 3. Determine whether observed loss rate can account for trapped flux evolution or whether other loss mechanisms are operational. 4. Confirm presence of EMIC waves (if they have propagated to low altitude) consistent with loss-cone spectra (baseline). 5. 2 CubeSats, <15min apart. | IR1. EPD–E range: 0.5-4MeV, δE/E≤5 of , pitch angle <28°. IR2. EPDE flux: Measure precipitating flux with sufficiently high sensitivity and low noise $10^2$-5·$10^5$#/$cm^2$·sr·s; or (for g-factor ~0.1 $cm^2$·sr) 10-50,000 counts/s IR3. Spectral cadence: Full angle/energy spectrum at every δL=0.5 from L=3 to L=5 (minimum) or δL=0.25 from L=2 to L=9 (baseline) once per ~90min. IR4. MAG: Determine DC pointing to within <2° i.e., resolution <500nT (minimum) and measure EMIC waves (baseline) at >2Hz rate, <1nT resolution. | MR1. Orbit: 350-6000km for science; <2500km for radiation; inclination, i >65°, arbitrary right ascension acceptable (minimum) with a 400km, circular, i=96°, sun-sync design as a baseline. MR1*. Orbit: i >65°, dawn-dusk ±75° MR2. Orbit control: Spin-up to 10-30 RPM, maintain spin plane <±15° off B-field for one auroral crossing per orbit (minimum); or 20RPM, <±15° from B once / orbit MR3. Duration: Reasonable precession of >15°/mo and storm occurrence (1 storm/mo.) give a 2 month residence in dawn-dusk orbits (optimal for EMIC studies) in a period <3mo (one-storm/month minimum) or more storms in a 6 month baseline period. |
| **Secondary** | Determine location of FAC sources relative to plasma boundaries (dipole vs. tail; inner edge of plasma sheet vs. plasma sheet boundary). | Measure latitudinal location of isotropy boundary (IB) of 50-300keV ions and 0.1-1MeV electrons with dMLAT~0.5° and compare with latitude of FACs to infer magnetospheric location of FAC sources relative to magnetospheric boundaries. STAR: 2 CubeSats | IR5. EPDI: Measure 50-300keV ions at ΔE/E≤50%, pitch-angle resolution <45°. IR6. EPDI flux: Measure 50-300keV ion flux above $10^3$ #/$cm^2$·sr·s, or (with g-factor >5.e-3 $cm^2$·sr) a count rate >5c/s. IR7. EPDE: Measure 50-500keV electron flux above 5·$10^5$ #/$cm^2$·sr·s (at >500 keV, IR2 is sufficient here too). IR8. MAG: δB<5nT sensitivity/stability, >1sample/s, DC-1Hz. | MR4. Activity: Assuming MR1 and MR2 are met, study non-storm time phenomena. |
| | Table 1. ELFIN minimum (red) and baseline (green) science objectives and traceability to instrument/mission requirements. | | | |

| Launch (L) | 2018-Sep-15 1:02UTC | |
|---|---|---|
| Release | ~ L + 1h 16min | |
| Satellite | ELFIN-A | ELFIN-B |
| Mean Altitude | 449.405 km | 449.446 km |
| Inclination | 93.031 deg | 93.0261 deg |
| Eccentricity | 0.00160 | 0.00157 |
| Right Ascension | 249.833 deg | 249.824 deg |
| Arg. of Perigee | 269.110 deg | 295.702 deg |
| Period | 93.735 min | 93.736 min |
| ***Table 2.** ELFIN launch orbit elements* | | |

| | Quantity | Requirement | Capability |
|---|---|---|---|
| **FGM** | Stability | 5nT / 2hr orbit | <1nT / 2hr orbit |
| | Sensitivity | <1nT | <0.1nT |
| | Resolution | <1nT | <0.1nT |
| | Noise | <1nT | <0.3nT (in-flight) |
| | $f_{Nyq., max}$ | >2Hz | >5Hz nominal |
| **EPD-E** | Energy range | 0.5 - 4 MeV | 0.05 - 5MeV |
| | $\delta E/E$ | <50% | ✓ (most < 40%) |
| | g-factor | >0.1 cm$^2$ sr | >0.13 cm$^2$ sr |
| | number flux | 5x10$^5$ /cm$^2$ s sr | ✓ |
| | FOV | <28$^o$ | 20$^o$ x 20$^o$ |
| **EPD-I** | Energy range | 50-300 keV | $^\dagger$250 - 5000 keV |
| | $\delta E/E$ | <50% | ✓ (most < 40%) |
| | g-factor | >5e-3 cm$^2$ sr | ~6.5 x 10$^{-3}$cm$^2$ sr |
| | number flux | 5x10$^5$ /cm$^2$ s sr | ✓ |
| | FOV | <28$^o$ | 22$^o$ x 22$^o$ |
| **Mission** | Inclination | >65$^o$ | 93$^o$ |
| | Altitude [km] | 350-2500 | ~450km (Launch) |
| | Radiation | 20kRads in 2mm, RDM2 for 6months | met through parts selection, spot shielding |

$^\dagger E_{ion,min}$=250keV due to out-of-spec dead layer

**Table 3.** *Instrument and mission baseline requirements versus demonstrated capability.*

| ELFIN Item | Mass [g] |
|---|---|
| FGM+FGE+harness | 257 |
| FGM boom | 450 |
| EPD mechanical | 769 |
| Bus Structures | 673 |
| Thermal | 45 |
| Power | 739 |
| Avionics Stack | 139 |
| Communications | 133 |
| ADCS coils | 159 |
| EPD+instr. electrical | 236 |
| **Total at launch** | **3,599** |
| Available (P-POD) | **4,000** |
| **Margin @ launch:** | **11%** |

**Table 4.** *Mass budget of ELFIN at launch.*

| ELFIN Subsystem | CBE + Cont. [W] | β=0 (no sci. zone) ON [min] | [W]/orbit | β=15 (1 sci. zone) ON [min] | [W]/orbit | β=30 (>4 sci. zones) ON [min] | [W]/orbit |
|---|---|---|---|---|---|---|---|
| Idle | 0.96 | 93.0 | 0.96 | 93.0 | 0.96 | 93.0 | 0.96 |
| Radio TX | 4.73 | 0.00 | 0.00 | 10.00 | 0.51 | 10.00 | 0.51 |
| Radio Beacon | 4.73 | 0.13 | 0.01 | 0.25 | 0.01 | 0.25 | 0.01 |
| Spin | 0.64 | 1.00 | 0.01 | 0.89 | 0.01 | 0.89 | 0.01 |
| Reor | 0.59 | 1.00 | 0.01 | 0.89 | 0.01 | 0.89 | 0.01 |
| Battery Heaters | 0.78 | 13.00 | 0.11 | 0.00 | 0.00 | 0.00 | 0.00 |
| IDPU | 0.46 | 0.00 | 0.00 | 8.00 | 0.04 | 20.00 | 0.10 |
| EPD only | 2.00 | 0.00 | 0.00 | 0.00 | 0.00 | 0.00 | 0.00 |
| FGM only | 0.89 | 0.00 | 0.00 | 2.00 | 0.02 | 8.00 | 0.08 |
| EPD+FGM | 2.89 | 0.00 | 0.00 | 5.00 | 0.16 | 20.00 | 0.62 |
| Total Draw | | | 1.09 | | 1.71 | | 2.29 |
| Total Available | | | 1.15 | | 1.75 | | 2.42 |
| Margin available | | | 5.6% | | 2.6% | | 5.7% |

**Table 5.** *Power budget of ELFIN based on pre-ship dat, for nominal attitude (orbit normal spin axis) and for spin-axis-to-sun angle β=0º (noon-midnight meridian orbit, limit cold-case no-science), β=15º (start of science orbits, more than one science zone of 5min duration per orbit possible), and β=30º (more than four science zones of 5min duration each possible per orbit).*

| ELFIN Downlink Budget | | | ELFIN Uplink Budget | | |
|---|---|---|---|---|---|
| Parameter | Value | Units | Parameter | Value | Units |
| *Spacecraft* | | | *Ground Station* | | |
| Carrier Frequency | 437.45 | MHz | Carrier Frequency | 145.96 | MHz |
| Wavelength | 0.69 | m | Wavelength | 2.05 | m |
| Tx Power Output | 1.80 | W | Tx Power Output | 100.00 | W |
| | 2.55 | dBW | | 20.00 | dBW |
| Peak Transmitter Antenna Gain | 0.42 | dBiC | Peak Transmitter Antenna Gain | 9.20 | dBiC |
| Transmit Antenna Pointing Loss | -12.80 | dB | Transmit Antenna Pointing Loss | -10.00 | dB |
| Line/Cable Losses: | -1.45 | dB | Line/Cable Losses: | -1.75 | dB |
| Spacecraft EIRP | 1.52 | dBW | Ground Station EIRP | 27.45 | dBW |
| *Path* | | | *Path* | | |
| Space Loss | -149.5 | dB | Space Loss | -139.9 | dB |
| Environmental Loss | -1.10 | dB | Environmental Loss | -0.80 | dB |
| Polarization Loss | -3.00 | dB | Polarization Loss | -3.00 | dB |
| *Ground Station* | | | *Spacecraft* | | |
| Receive Antenna Pointing Loss | -3.00 | dB | Receive Antenna Pointing Loss | -10.80 | dB |
| Isotropic Signal Level at GS | -167.85 | dBW | Isotropic Signal Level at SC | -136.29 | dBW |
| Peak Receive Transmitter Antenna Gain | 45.00 | dBi | Peak Receive Transmitter Antenna Gain | 3.10 | dBi |
| Effective System Noise Temperature | 1717 | K | Effective System Noise Temperature | 615 | K |
| GS System Figure of Merit (G/T): | 12.65 | dB/K | SC System Figure of Merit (G/T): | -24.79 | dB/K |
| Ground Station C/No | 73.40 | dBHz | Spacecraft C/No | 67.52 | dBHz |
| *Data* | | | *Data* | | |
| Data Rate | 19200 | bps | Data Rate | 9600 | |
| | 42.83 | dBHz | | 39.82 | dBHz |
| Bit Error Rate | 1E-05 | | Bit Error Rate | 1E-05 | |
| Calculated Eb/N0 | 30.57 | dB | Calculated Eb/N0 | 27.70 | dB |
| Expected Eb/N0 | 15.80 | dB | Expected Eb/N0 | 13.80 | dB |
| **Link Margin** | **14.77** | **dB** | **Link Margin** | **13.90** | **dB** |

**Table 6.** *ELFIN's downlink (left) and uplink (right) budgets.*

| Operational Scenario | Sci. Zone Types / Orbit | MB/day [x2 Compression] | Incl. HSK [MB/day] |
|---|---|---|---|
| Storm | 4 Fast | 0.665 | 0.99 |
| Normal | 1 Fast, 3 Slow | 0.215 | 0.54 |
| Minimum | 1 Fast | 0.04 | 0.06 |

**Table 7.** *ELFIN's data volume.*

**Figures**

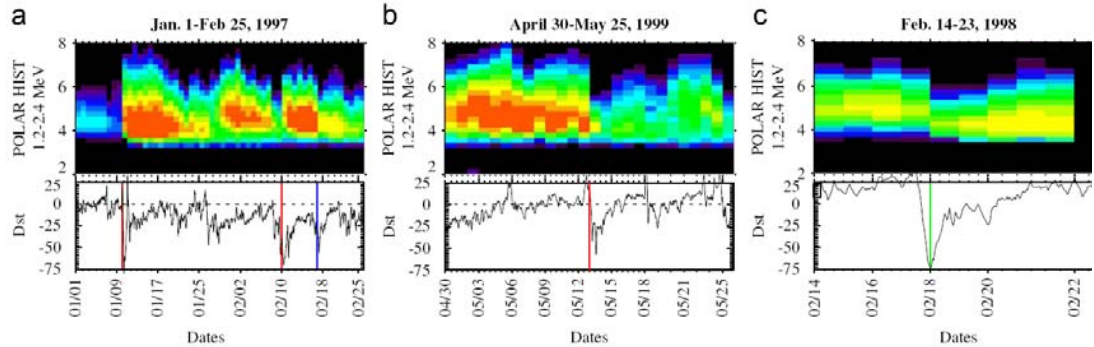

***Figure 1.*** *Details of the three types of radiation belt responses to storm-time ring current enhancements. (a) A strong increase in relativistic electron fluxes in response to the January 1997 geomagnetic storm. (b) A dramatic and permanent loss of electrons throughout the outer belt in May 1999. (c) A 100 nT storm in February 1998 with post-storm fluxes similar to pre-storm.*

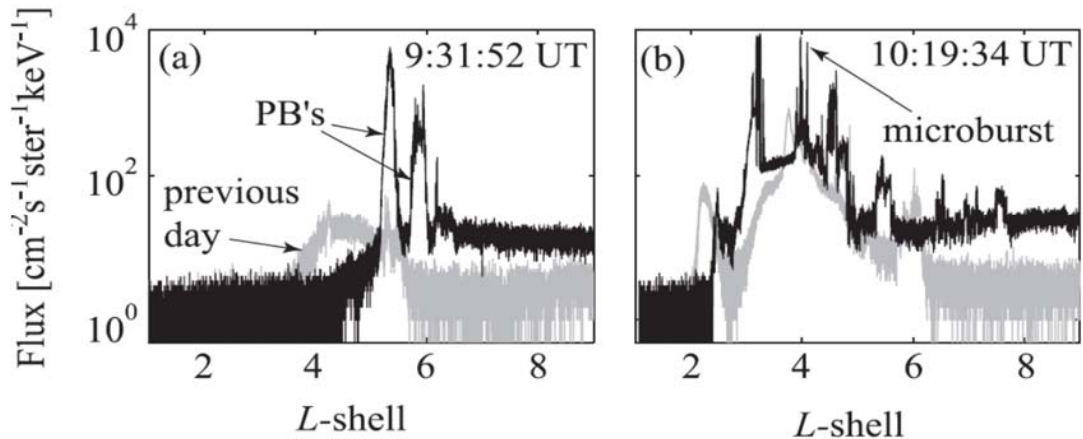

***Figure 2.*** *SAMPEX data (from Bortnik et al., 2006) showing examples of precipitation bands (PBs) on 20 November 2003 at (a) the beginning of the storm and (b) during the storm main phase. The SAMPEX flux at the same location, but on the previous day (19 November 2003), is shown in gray.*

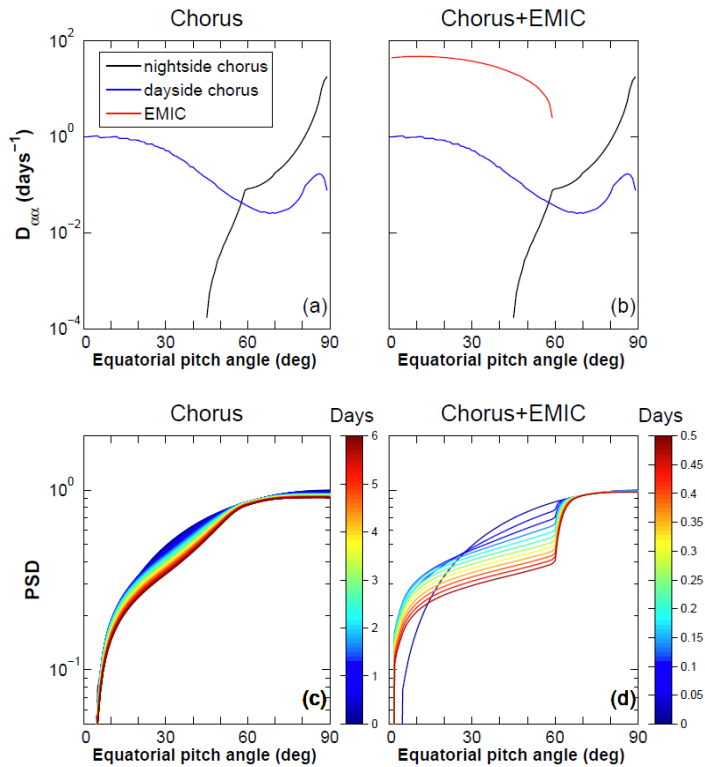

**Figure 3.** *Evolution of normalized 1MeV electron phase space density as a function of pitch angle at L=4.5 using the wave-scattering model in Li et al. (2007) Top: Bounce-averaged pitch-angle scattering rates caused by chorus only (left) and chorus plus EMIC waves (right). Bottom: Evolution of phase space density (color coded by time in days) due to chorus only (left) and chorus plus EMIC waves (right). EMIC waves are even more effective at energies >1MeV.*

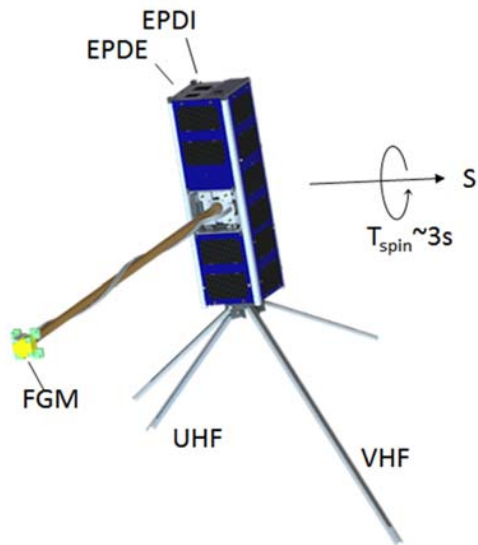

**Figure 4.** *ELFIN with its deployed fluxgate magnetometer boom and spin axis, **S**, which will be nominally positioned near the orbit normal.*

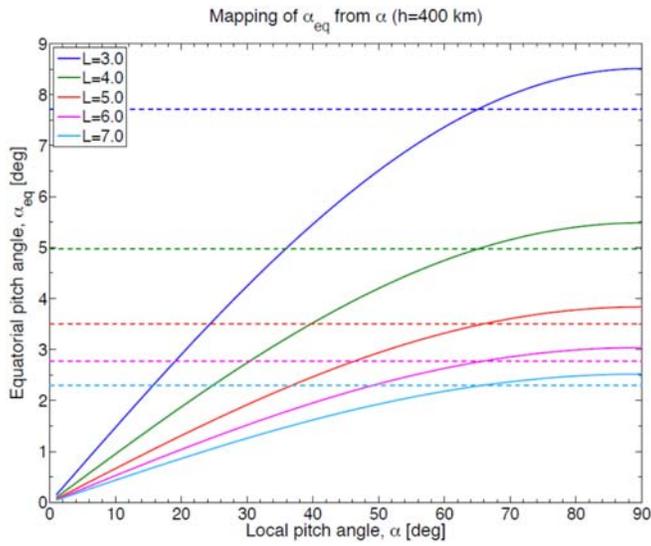

**Figure 5.** *Equatorial versus local pitch angle (at the anticipated life-time average ELFIN altitude of h=400km, at various L-shells (in color). Dashed horizontal lines indicate the equatorial loss cone.*

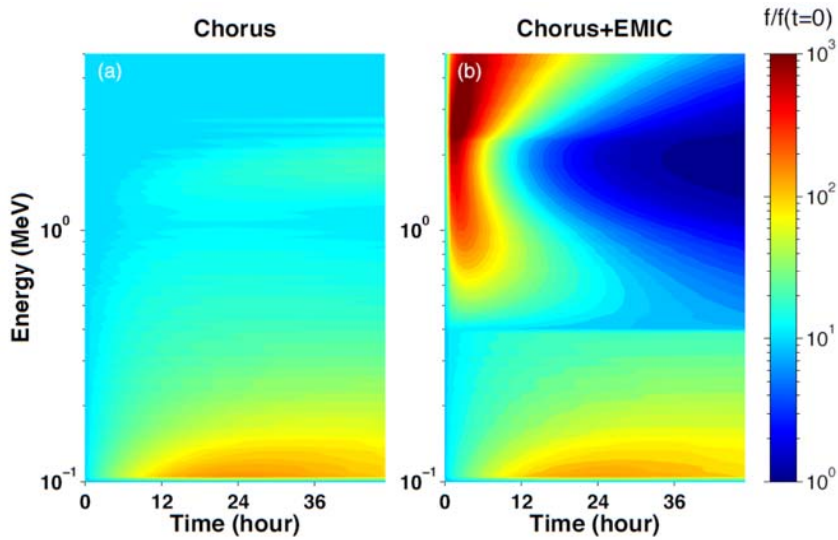

**Figure 6.** *Phase space density (f) evolution near the loss cone (normalized to f at t=0), using the wave model of Li et al. 2007; and Shprits et al., 2008 and assuming scattering by chorus only (left) and chorus plus EMIC waves (right) at L=4.5.*

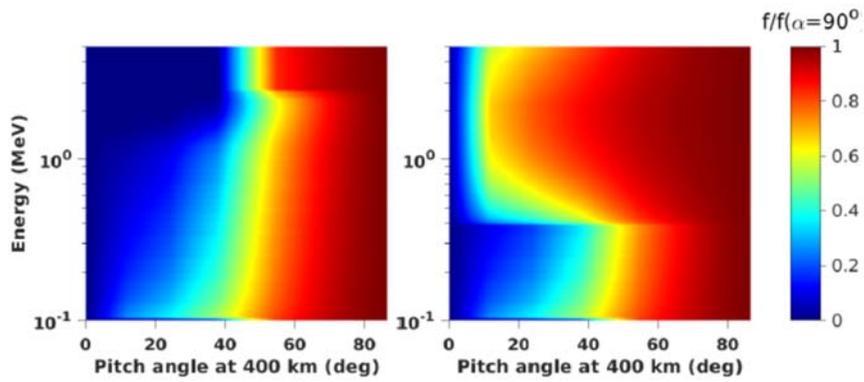

**Figure 7.** *Energy vs. pitch-angle spectra, at 400km (L=4.5) normalized by the trapped flux (at local pitch angle α=90° ) after 12 hrs of interaction, using the wave model in Figure 6 (Li et al., 2007, Shprits et al., 2008).*

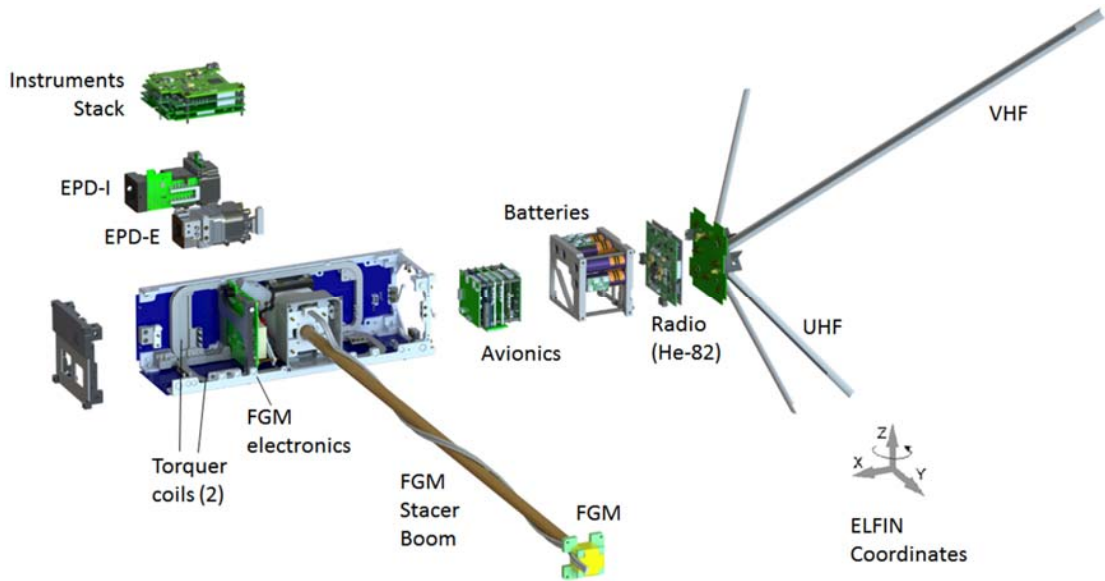

**Figure 8.** *ELFIN exploded view after boom deployment.*

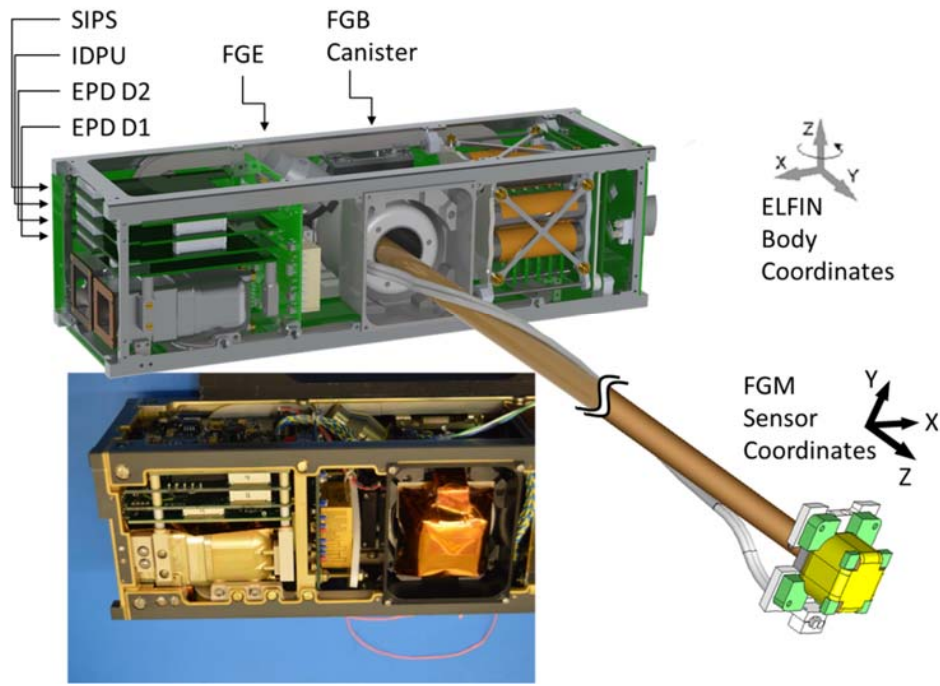

***Figure 9.*** *ELFIN instruments in ELFIN's 3U volume in CAD fashion (with magnetometer boom shown deployed) and as an image after instrument integration into the spacecraft. The instrument stack boards shown atop the EPD are the Switching Instrument Power Supply (receiving raw power and conditioning and distributing it to the two instruments), the Instrument Data Processing Unit (discussed in the text), and two digital processing EPD boards, D1 and D2, which interface with the IDPU on the one side and the Electronics Front End (EFE) and Preamplifier (Preamp) boards of the EPD instrument on the other. The latter two boards are not visible here but will be discussed in the EPD paper. Additionally the Fluxgate Magnetometer Electronics (FGE) board is shown in the schematic and after mission integration. It hosts all electronics needed to operate the FGM and interfaces directly with the IDPU. Nearly all of the FGM electronics have been integrated into the Multi-Chip Module (gold, most sizeable component). To avoid drum-heading, a Delrin strap (visible in the picture on top of the MRM) was attached to an H-brace at the back of the FGE late in the program. The FGM sensor was blanketed prior to flight (orange blankets evident in the image were not included in the CAD schematic). The FGM stacer boom (FGB), coiled inside a canister in the stowed configuration, was deployed by the tension release initiated by a shaped memory alloy actuator, a lug attached to (but external to) the FGB, visible in both the schematic and the picture between FGE and FGB.*

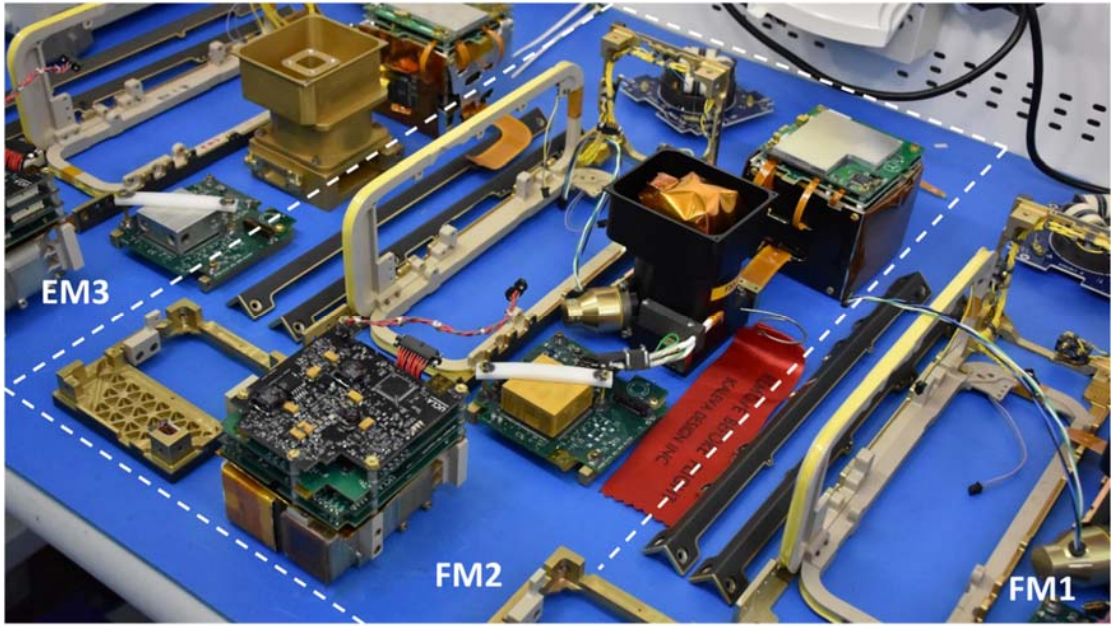

**Figure 10.** *A table-top view of all the satellite subsystems, including instruments, bus avionics, torquecoils and structure, is seen prior to spacecraft integration and environmental tests. The picture zooms into the Flight Model 2 (FM2) components (encompassed by the white dashed parallelogram). Components for FM1 and the highest fidelity engineering model (EM3) are also depicted on its two sides.*

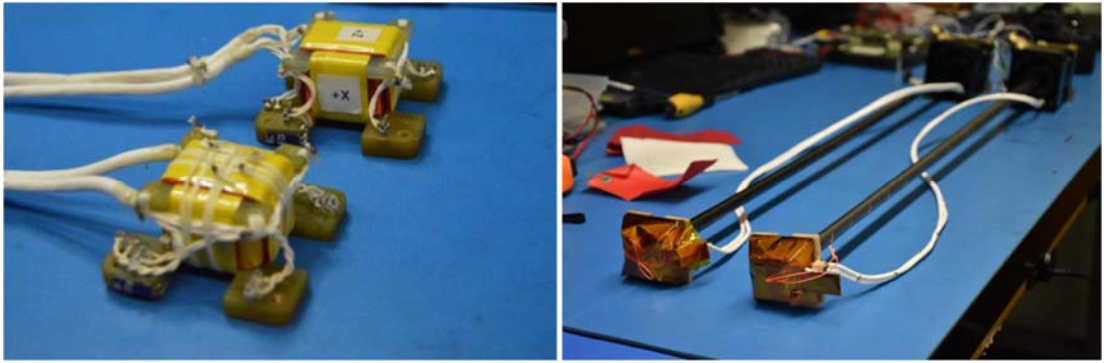

***Figure 11.*** *ELFIN's flight FGM sensors before integration to the spacecraft (left) and after FGM flight-unit boom deployment tests (right).*

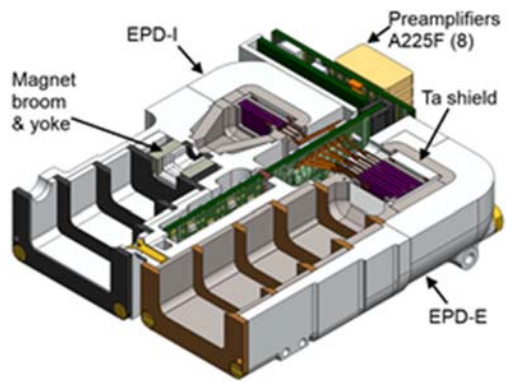

***Figure 12.*** *Schematic of ELFIN's EPD detector comprising an ion and electron head. The bias and front-end electronics modules are in a board sandwiched between the two heads; while the preamp board is behind the shorter ion head.*

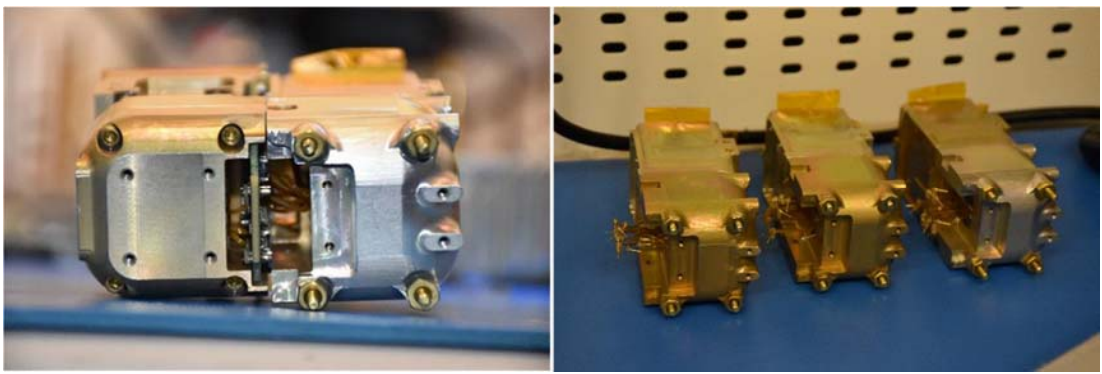

***Figure 13.*** *Pictures of integrated EPD sensor heads: Left: view (from back) of both ion and electron heads (+Z is down, electron head is on the right), showing the front-end electronics module. Right: two FMs and one EM of electron detector heads after integration with detectors.*

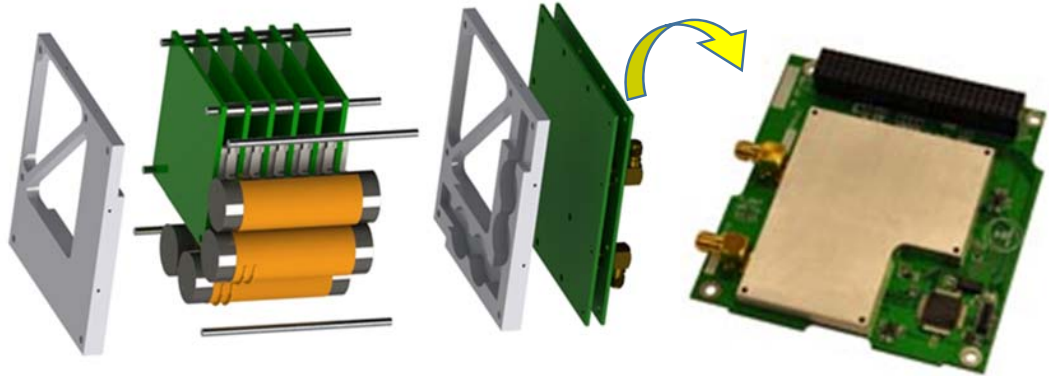

**Figure 14.** *Bus components on the avionics side. The stack contains six avionics boards and four batteries provided by The Aerospace Corporation, plus two boards: the Big etcetera (BETC) board, and the radio board, a "Helium-82" radio by AstroDev (VHF up/UHF down). A top-down view of the radio is shown on the right. The stack is encapsulated in shielding material and blankets to ensure good thermal isolation and control of the batteries, and electromagnetic noise isolation from the instruments. The integrated stack is shown prior to bus integration in Fig. 10.*

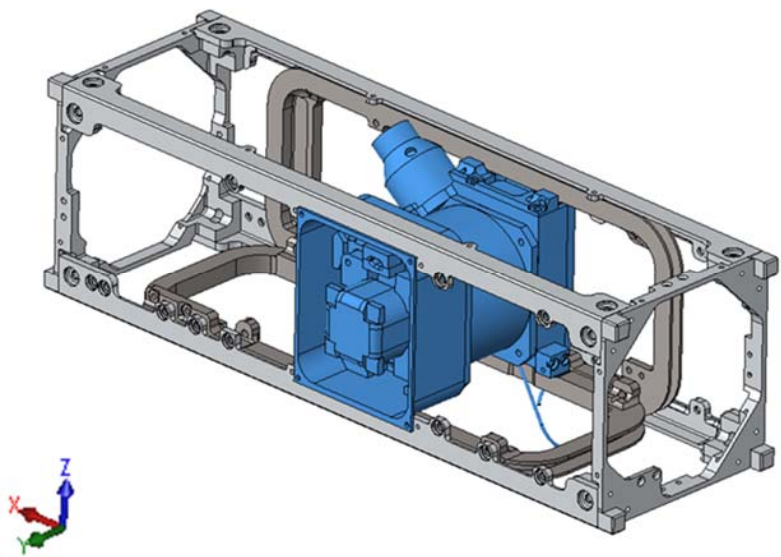

**Figure 15.** *Bus mechanical structure.*

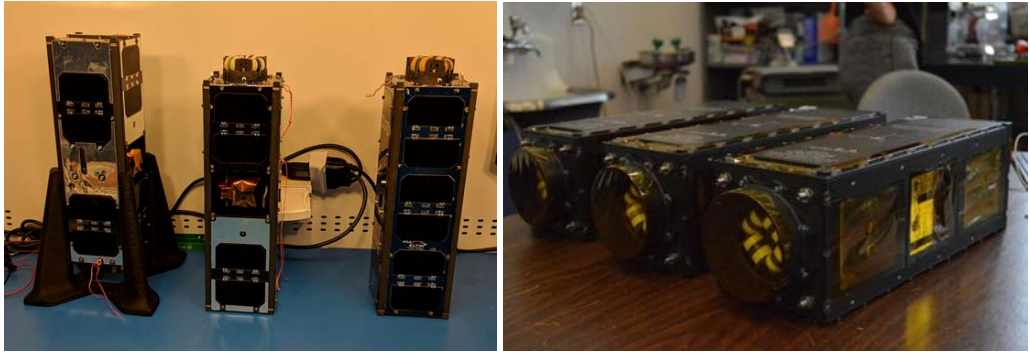

**Figure 16.** Left: Integrated ELFIN spacecraft (from left to right: two FMs and one EM). Right: Spacecraft inside PPODS, showing how antenna tuna-can takes advantage of the extra volume available for launch.

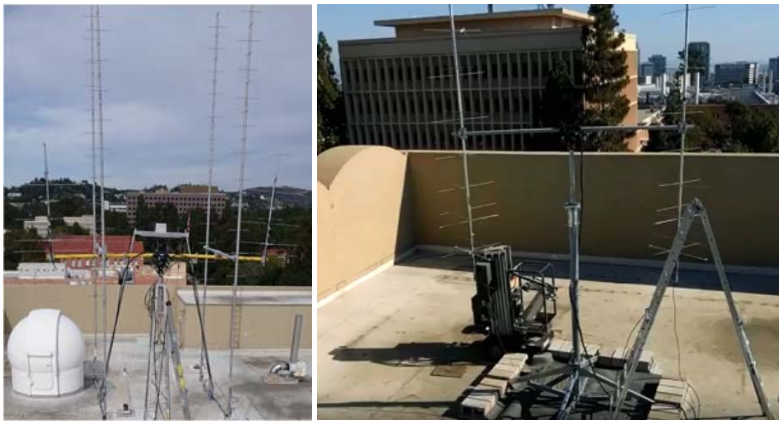

*Figure 17. Left: UCLA's Knudsen North station with four UHF yagi antennas (downlink) and two VHF yagis (uplink, using power 10W). The spacing between UHF antennas is 6 feet on each side, and the spacing between VHF antennas is 10 feet. Right: Knudsen South station is a two VHF yagi providing higher output power (100W). Both stations can be used simultaneously and provide flexibility when both ELFIN satellites are within view.*

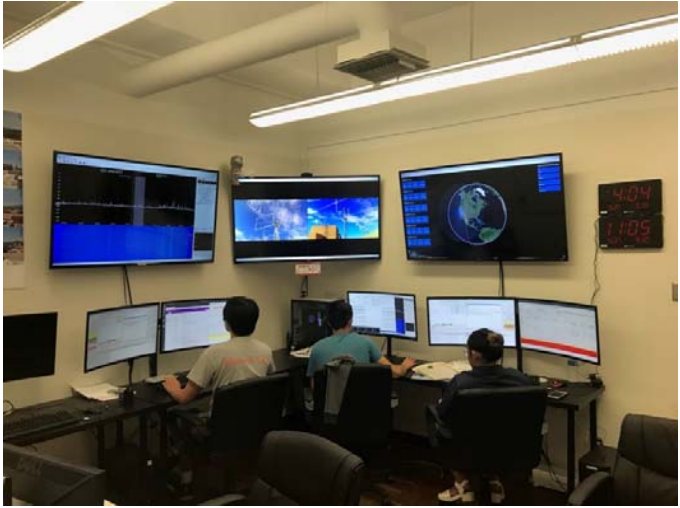

**Figure 18.** *ELFIN's Mission Operations Center at UCLA.*

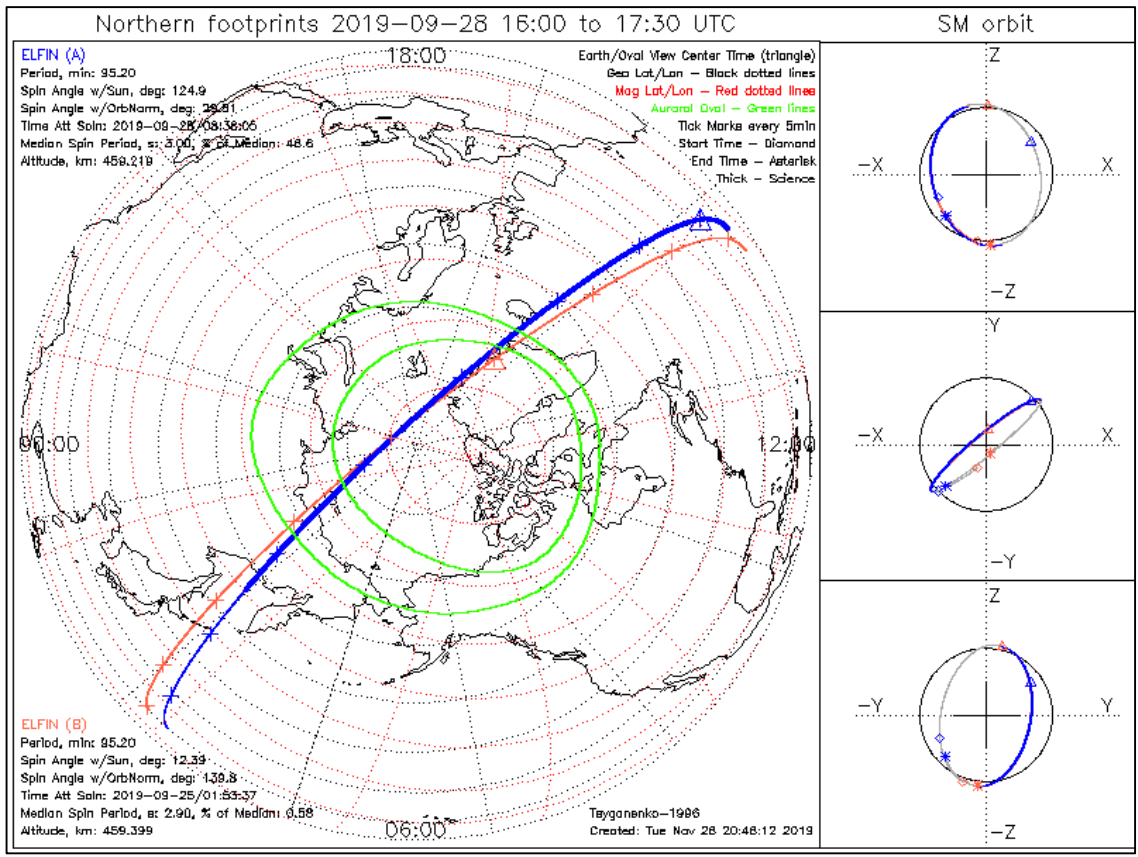

**Figure 19.** *ELFIN's orbit plots (example taken from: https://elfin.igpp.ucla.edu/ → Science Overview → Summary Plots) depict definitive position (orbit projected at 100km) and attitude (tabulated as an insert at the top left and bottom left of the plot), as well as future position information, allowing scientific event selection and future science operations planning of conjunctions with other missions. This tool complements SPDF's TIPSOD program (a 4D orbit viewer) that is applied on the same orbit dataset, released to SPDF daily.*

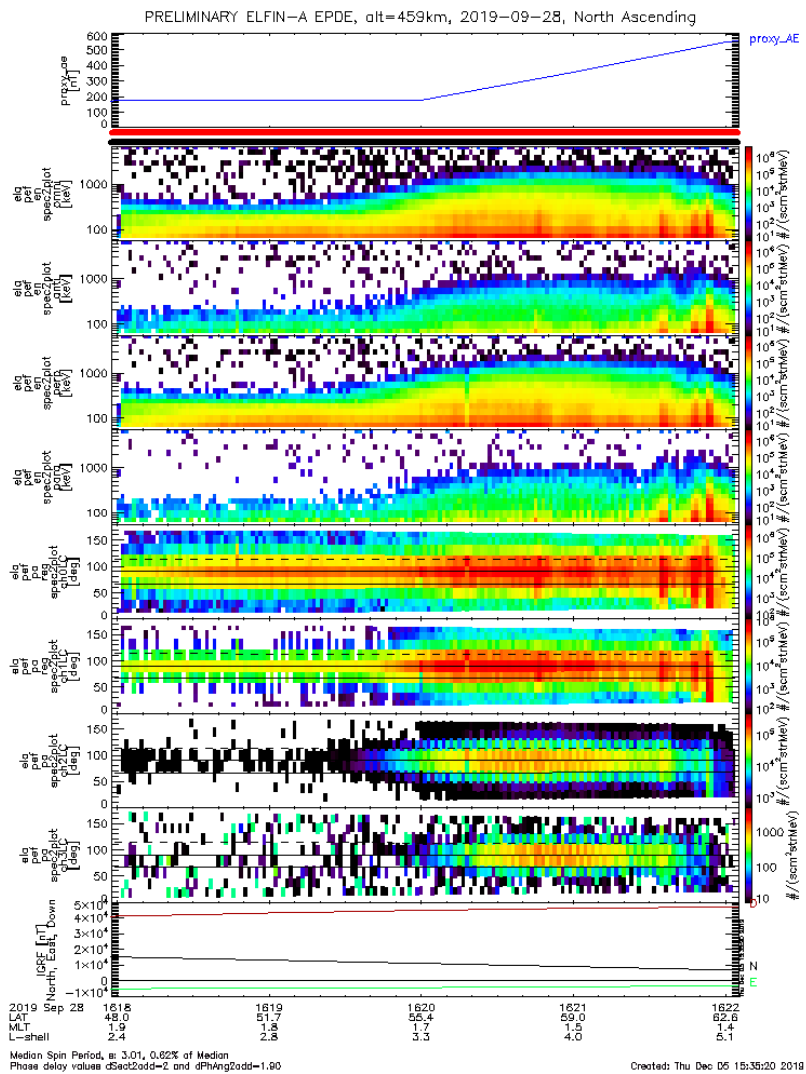

***Figure 20.*** *ELFIN's summary plots routinely produced by science operations depict the primary dataset (EPDE energy and pitch-angle spectrograms) along with information on loss cone and field direction. Shown is an example of a science zone crossing (ascending North, post-midnight sector). A key on the web-site provides detailed information on the various panels. The top four spectrograms are energy spectrograms of the electron number flux (omni-directional, anti-parallel, perpendicular and parallel to the magnetic field, respectively). The bottom four spectrograms are pitch-angle spectrograms of electron energies [[50.,160.],[160.,345.],[345.,900.],[900.,7000.]]. The pitch-angle spectrograms also contain a solid line at 90º, and a second solid line demarcating the loss cone based on the IGRF magnetic field given the spacecraft location and attitude. The dashed line is the anti-loss cone direction (complement of the loss cone). The bottom panel is the IGRF magnetic field in local magnetic coordinates: North (horizontal), East, and Down (vertical).*

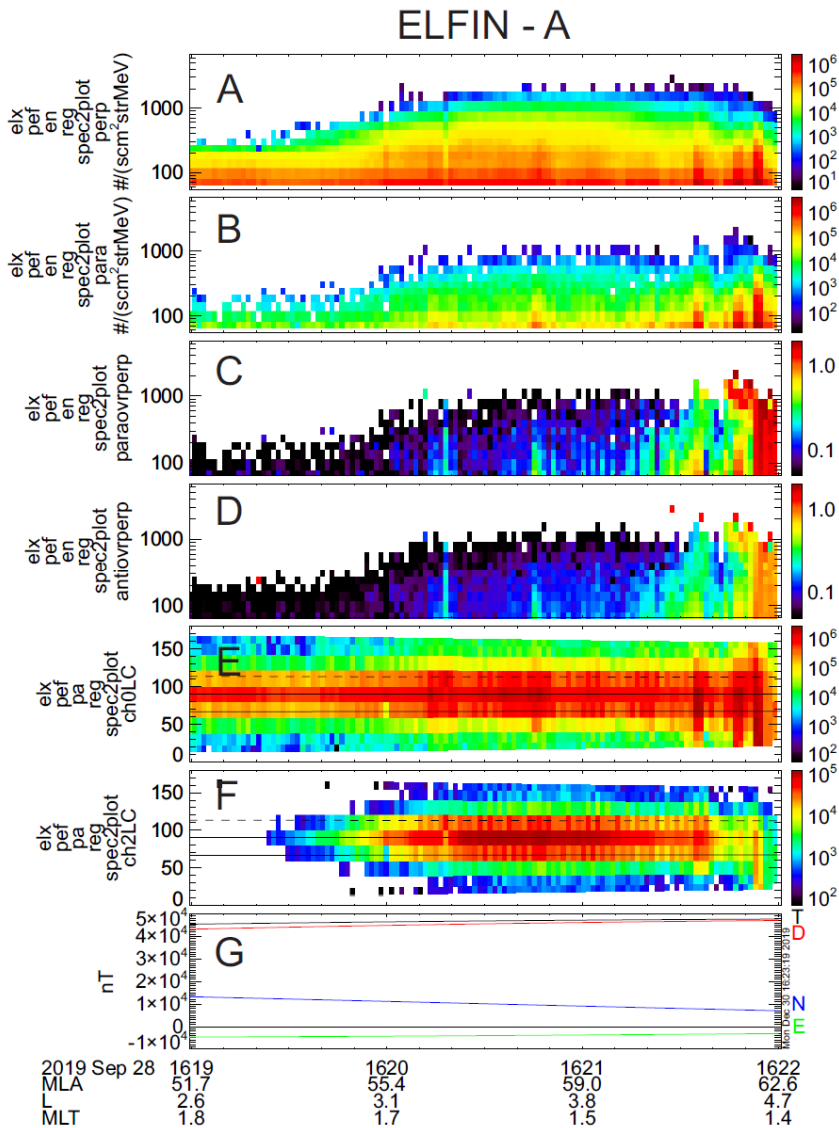

**Figure 21.** *ELFIN EPDE plot produced using nominal data analysis by the ELFIN software (a SPEDAS plug-in). The analysis tools will be released together with the data once routine collections commence with sufficient volume through the ELFIN website. The event shown, a subset of Fig. 20, demonstrates the scientific potential of the mission. A: energy spectrogram of perpendicular electron fluxes (trapped). B: energy spectrogram of parallel (down-going or precipitating) electron fluxes. C: ratio of parallel to perpendicular fluxes. D: ratio of anti-parallel (upward-moving or backscattered) fluxes. E: pitch-angle spectrogram of 50-160 keV electron number flux (in #/(s cm² str MeV)); F: same as E but for 345-900keV; G: Earth's IGRF model magnetic field in local magnetic coordinates: North (horizontal magnetic North), East (horizontal magnetic East), and Down (vertical).*